\documentclass[fleqn,usenatbib]{mnras}

\usepackage{newtxtext,newtxmath}

\usepackage[T1]{fontenc}

\DeclareRobustCommand{\VAN}[3]{#2}

\let\VANthebibliography\thebibliography
\def\thebibliography{\DeclareRobustCommand{\VAN}[3]{##3}\VANthebibliography}

\usepackage{xcolor}
\usepackage[normalem]{ulem}
\usepackage{booktabs} 
\usepackage{siunitx}
\definecolor{highlight}{RGB}{217,238,255} 
\usepackage{natbib}

\usepackage{graphicx}	
\usepackage{amsmath}	
\usepackage{subcaption}
\usepackage{lastpage}
\usepackage{verbatim}
\usepackage{caption}

\newcommand\sersic{S\'ersic}
\newcommand\reff{$R_{\rm eff}$}
\newcommand\galight{\texttt{galight}}

\newcommand\mum{\textmu m}

\newcommand{\red}[1]{\textcolor{black}{{#1}}}
\newcommand{\green}[1]{\textcolor{black}{{#1}}}

\newcommand\coloredsout[2][red]{\bgroup\markoverwith
{\textcolor{#1}{\rule[0.5ex]{2pt}{0.4pt}}}\ULon{#2}}

\title[Resolving Galaxy Bulge+Disk Structure with JWST]{Unveiling Galaxy Structures: Systematic Analysis of Bulge+Disk Decomposition Using Simulated JWST/NIRCam Observations}

\author[Tang et al.]{\parbox{\textwidth}{
Zhengxin Tang,$^{1}$
Xuheng Ding$^{1}$\thanks{E-mail: dingxh@whu.edu.cn}
Boris S. Kalita,$^{2,3,4}$
Lilan Yang,$^{5}$ 
}
\\
\\
\parbox{\textwidth}{
$^{1}$School of Physics and Technology, Wuhan University, Wuhan 430072, China\\
$^{2}$Kavli IPMU (WPI), UTIAS, The University of Tokyo, Kashiwa, Chiba 277-8583, Japan\\
$^{3}$Kavli Institute for Astronomy and Astrophysics, Peking University, Beijing 100871, People's Republic of China\\
$^{4}$Centre for Data-Driven Discovery, Kavli IPMU (WPI), UTIAS, The University of Tokyo, Kashiwa, Chiba 277-8583, Japan\\
$^{5}$Laboratory for Multiwavelength Astrophysics, School of Physics and Astronomy, Rochester Institute of Technology, 84 Lomb Memorial Drive, Rochester, NY 14623, USA
}}

\date{Accepted XXX. Received YYY; in original form ZZZ}

\pubyear{\the\year{}}

\begin{document}
\label{firstpage}
\pagerange{\pageref{firstpage}--\pageref{lastpage}}
\maketitle

\begin{abstract}
Characterizing and accurately decomposing galaxies into structural components, such as bulges and disks, is essential for understanding galaxy formation and evolution, particularly at high redshift, where galaxies are compact and faint. Leveraging the unparalleled resolution and sensitivity of JWST and imaging data from CEERs program, we simulate galaxies with bulge+disk components and assess the effectiveness of single and double \sersic\ model fittings, respectively. 
We first evaluate the performance of single \sersic\ fits, and find it can recover total magnitudes (i.e., within 0.5 mag), and size (i.e., within 0.2 dex), down to 27 mag. 
The features that emerged in the residual map can properly reflect the underlying two-component structures. We also show that \sersic\ indices can serve as proxies for the bulge-to-total flux ratio (B/T). 
For double \sersic\ models, we find comparable accuracy in recovering bulge and disk magnitudes (i.e., within 0.5 mag), and effective radius (i.e., within 0.2 dex), down to 26 mag. 
To quantitatively determine whether a double \sersic\ model better describes our two-component systems compared to a single \sersic\ profile, we evaluate the Bayesian Information Criterion for both model configurations.
To extend the applicability of our results to other NIRCam programs, we evaluate the signal-to-noise ratio (SNR) of the simulated galaxies and find that model parameters are reliably reproduced when the SNR exceeds 10.
Our work demonstrates the detailed morphological measurement uncertainties using single and double \sersic\ models, which provides an essential reference for future JWST/NIRCam-based morphological studies, especially for high-redshift galaxies.
\end{abstract}



\section{Introduction}
Galaxy morphology serves as a critical medium for understanding the formation and evolution of galaxies~\citep[e.g.,][]{2004ARA&A..42..603K,2005ARA&A..43..581S,mo2010galaxy,2014ARA&A..52..291C}. The structural components of galaxies -- such as bulges, disks, and bars -- encode valuable information about their assembly history,  including mergers, accretion events, and secular evolution processes~\citep[e.g.,][]{2005MNRAS.358.1477A,2009MNRAS.393.1531G,kormendy2013secular,du2021evolutionary}. 
Accurate morphological decomposition, which separates these components, is essential for testing theoretical models of galaxy formation and provides insights into the physical mechanisms driving galaxy evolution~\citep[e.g.,][]{peng2002detailed,2011ApJS..196...11S,
2012MNRAS.421.2277L,2014ApJ...788...11L}.

Despite its importance, accurately decomposing galaxy structures remains a significant challenge, particularly for high-redshift galaxies~\citep[e.g.,][]{peng2002detailed,peng2010detailed,2012ApJS..203...24V,2013MNRAS.430..330H,bruce2014}. These galaxies are often more compact and have \green{dim surface brightness}~\citep[e.g.,][]{1996ApJ...470..189G,
2007MNRAS.382..109T,vanderWel2014}, making it difficult to distinguish between structural components.
A common practice to model high-redshift galaxies' light profiles is to use analytical models to describe the overall surface brightness distribution of galaxies. Among these, the \sersic\ profiles~\citep[e.g.][]{sersic1963} have been widely adopted as a first-order approximation for the light distribution of typical galaxies. The \sersic\ profile is flexible, allowing it to describe a range of galaxy types, from exponential disks ($n=1$) to de Vaucouleurs' laws for elliptical galaxies ($n=4$). While the \sersic\ profile is a powerful tool for describing galaxy light distributions, only a single-component form struggles to capture the composite structure of systems like bulge+disk galaxies. Fitting such systems with a single \sersic\ profile could introduce biases in derived parameters -- including luminosity, size, and morphological indices, due to the oversimplification of distinct kinematic and photometric components~\citep[e.g.,][]{2011ApJS..196...11S,2012MNRAS.421.2277L,2013MNRAS.433.1344M,2014MNRAS.443..874B}.

The launch of the JWST has opened a new era in the study of galaxy morphology at high redshift~\citep[e.g.,][]{Yang2022,Treu2023, 2022ApJ...939L..28D, montes2022new, sun2024structure}. With its unprecedented spatial resolution~\citep[$< 0\farcs1$ at 2~$\mu$m; e.g.,][]{2023PASP..135b8001R} and broad wavelength coverage from 0.6 to 28~$\mu$m, spanning both the near-infrared (NIRCam: 0.6-5.0~$\mu$m) and mid-infrared (MIRI: 5.0-28.0~$\mu$m) regions, JWST enables detailed structural analysis of distant, faint, and compact galaxies that were previously inaccessible. This capability has already led to remarkable discoveries of well-ordered disk galaxies and morphological diversity at $z \gtrsim 6$ \citep[e.g.,][]{2022ApJ...938L...2F}, indicating that structural maturity may emerge earlier than previously thought.

Despite these advances, the majority of morphological studies with JWST have thus far relied on single-component \sersic\ profile fitting to characterize global galaxy structure (e.g., effective radius, \sersic\ index), continuing a tradition established with  \textit{Hubble Space Telescope} (HST) and ground-based surveys~\citep[e.g.,][]{peng2002detailed, peng2010detailed, 2017MNRAS.466.2024T}. Recent JWST-based studies have successfully applied single \sersic\ fits to large samples at $z \gtrsim 2$~\citep[e.g.,][]{miller2024jwstuncoversopticalsize, blanchard2024jwst, LaChance_2025, Yang2025}, confirming the compact nature of early galaxies and enabling size evolution studies.

However, systematic studies of multi-component structural decomposition -- particularly bulge+disk modelling using double \sersic\ profiles -- remain scarce in the JWST literature. Two open questions arise: (1) Can JWST reliably distinguish between single- and multi-component systems at high redshift? (2) How accurately can the physical properties of individual components (e.g., magnitudes, sizes, B/T) be recovered under JWST-like conditions?

Most previous simulation-based validation efforts have largely focused on single-component fitting or relied on lower-resolution datasets from surveys such as SDSS and KiDS~\citep[e.g.,][]{meert2013simulations, mosleh2013robustness, casura2022galaxy}. In parallel, early attempts to simulate high-redshift galaxies for morphological analyses were carried out using HST observations~\citep[e.g.,][]{2004ApJ...615L.105J}, highlighting the inherent difficulties of applying structural decomposition techniques at early cosmic epochs. While these studies have been instrumental in developing robust fitting pipelines, they do not address the new challenges posed by high-resolution, high-redshift imaging enabled by JWST. In particular, the lack of rigorous, JWST-specific validation for two-component (i.e., double \sersic) modelling represents a key gap in current methodology.

Our study aims to fill this gap by systematically assessing the performance and limitations of bulge+disk decompositions under JWST-like conditions. Such validation is critical for obtaining reliable morphological measurements and for interpreting the physical processes driving galaxy assembly in the early Universe. We focus on JWST's primary imager, the Near Infrared Camera (NIRCam), which delivers high-quality imaging data for galaxies across a wavelength range of 0.6 to 5~\mum. We generate mock bulge+disk galaxies and embed them into NIRCam observational data, incorporating a realistic point spread function (PSF) and background noise based on JWST. To evaluate the performance of different modelling approaches, we first fit each mock galaxy with a single \sersic\ profile. This allows us to assess whether the fitting outputs (such as residual maps) can reveal the underlying two-component (bulge+disk) structure. We then apply a double \sersic\ model to decompose the galaxies into their constituent bulge and disk components, quantifying the accuracy of recovered properties (e.g., magnitudes, effective radius, and B/T). By systematically quantifying biases and uncertainties across a wide range of parameter space -- including magnitude, B/T, and SNR -- we establish empirical error budgets and practical constraints for future JWST observations.

This paper is organized as follows. Section~\ref{observe_data} details our materials from the CEERS observation, including PSF preparation and noise incorporation. Section~\ref{sec.3} presents our bulge+disk simulation methodology and fitting procedures. In Section~\ref{sec.4}, we apply the simulation to fit with single and double \sersic\ models, quantifying parameter recovery accuracy, SNR-dependent biases, and model selection via the BIC. Section~\ref{sec.5} discusses the utility of single \sersic\ indices for estimating B/T, validates results against \textsc{galfit}, and justifies our focus on bulge+disk decompositions. We conclude in Section~\ref{sec.6} with implications for galaxy evolution studies. Throughout, we adopt AB magnitudes and a flat $\Lambda$CDM cosmology with $\Omega_{m}=0.3$, $\Omega_{\Lambda}=0.7$, and $H_{0}=70\,\mathrm{km\,s^{-1}\,Mpc^{-1}}$.



\section{Materials from CEERS observation}\label{observe_data}

We aim to create simulations that closely mimic the observational capabilities of the JWST.  To this end, our simulations are based on publicly available data from the CEERS program, a widely recognized and publicly available survey designed to explore galaxy populations in the early universe.

The CEERS program (ERS Program \#1345, PI: Finkelstein) is designed to explore approximately 100~square arcminutes of the AEGIS field through extensive imaging and spectroscopic observations~\citep[e.g.,][]{Finkelstein2017,Finkelstein2022}. The CEERS campaign features 10~NIRCam pointings, complemented by 6~NIRSpec and 4~MIRI parallel pointings, providing a rich multi-wavelength view of early galaxy populations. Observations utilized seven filters (F115W, F150W, F200W, F277W, F356W, F410W, and F444W), with integration times of approximately 2,635~seconds per filter, except for F115W, which received double the exposure time.  The 5$\sigma$ point-source detection depth of CEERS under a 10-hour exposure is approximately AB 29 mag~\citep[e.g.][]{Finkelstein2022}. Details on the JWST observations, including dither strategies, data reduction steps, and initial scientific results, can be found in the published literature~\citep[e.g.,][]{Finkelstein2022,Kelly2023}. 
The data were processed using the JWST Calibration Pipeline (version~1.8.5), which includes standard bias, dark, flat-field, and flux calibration steps~\citep[e.g.,][]{Rigby2022}. These calibrations ensure that the final science images meet the quality requirements needed for quantitative morphological analysis. The final pixel scale of the data production is 0\farcs03 for all bands, which we will adopt as the pixel scale for our simulations.

To comprehensively sample NIRCam's wavelength range while avoiding redundancy, we select two representative filters from the seven available in the CEERS program: F150W (\SI{1.5}{\micron}) in the short-wavelength (SW: \SIrange{0.6}{2.3}{\micron}) channel and F356W (\SI{3.56}{\micron}) in the long-wavelength (LW: \SIrange{2.4}{5.0}{\micron}) regime.  The F150W filter lies near the midpoint of the SW channel and offers superior angular resolution (<0$\farcs{}$05), making it ideal for resolving compact structures like bulges. Conversely, the F356W filter resides in the LW channel, where it benefits from reduced background noise and enhanced sensitivity to stellar light at high redshifts, as it is above the 4000~\AA\ break at $z$ up to 7.
By selecting both SW and LW regimes, the two filters capture NIRCam's dual capabilities: high spatial resolution for structural decomposition (F150W) and \green{longer wavelength coverage for probing faint and distant galaxies } (F356W). This strategic selection balances observational efficiency with the need to span NIRCam's full wavelength performance, ensuring our simulations reflect the instrument's versatility across key science cases.

Our simulations incorporate two critical observational elements from the CEERS survey: {\it background condition} and {\it PSF models}, which together replicate JWST/NIRCam's environmental and instrumental conditions. We specifically use the CEERS5 pointing -- located in the well-studied AEGIS field -- as the basis for our simulated observations.

{\it Background Condition}: \green{Our mock galaxies will be injected into random empty sky regions extracted from CEERS NIRCam F150W and F356W imaging data}. This preserves the noise properties (e.g., random white noise and detector artifacts) of actual JWST observations, ensuring our simulated galaxies experience realistic background fluctuations.

{\it PSF Models}: To account for spatial variations in the PSF across NIRCam's field of view, we use the \texttt{find\_psf} function from the \texttt{galight} package (Section~\ref{galight}) to identify PSF candidates. These candidates are bright, isolated stars with unsaturated profiles and stable full width at half maximum (FWHM) values, ensuring they accurately represent the instrument's optical response. After careful visual \green{inspections}, we selected four PSFs for F150W and five PSFs for F356W from CEERS5, chosen to span the range of observed PSF shapes and sizes in each band. These empirically derived PSFs capture wavelength-dependent diffraction patterns (e.g., \green{sharper cores in F150W and broader wings in F356W, respectively}) and spatial variations inherent to NIRCam, ensuring our simulations reflect actual on-sky performance during CEERS observations.

The detailed methodology for generating these mock galaxies is described in the following section.

\section{Simulation Generation and Fitting Methodology}\label{sec.3}

The \sersic\ profile is a widely adopted analytical model to represent galaxies' surface brightness distribution. In this study, our simulations and fittings rely on this profile. The \sersic\ model is defined as follows:
\begin{equation}
I(R) = I_e \exp\left[-k\left(\left(\frac{R}{R_{\text{eff}}}\right)^{1/n}-1\right)\right],
\label{sersic}
\end{equation}  
where $I_e$ is the surface brightness at the half-light radius $R_{\text{eff}}$, and $n$ is the \sersic\ index. The axis ratio and other parameters are free to vary. The parameter $k$ ensures 
$R_{\text{eff}}$ contains half the flux and depends on $n$. This model generalizes galaxy light profiles: $n=1$ corresponds to exponential disks; $n=4$ matches de Vaucouleurs' laws for elliptical galaxies.

This study adopts a two-stage methodology to evaluate galaxy modelling/decomposition techniques under JWST observational conditions. First, we generate mock bulge+disk galaxies incorporated with PSF models and background noise based on CEERS/NIRCam data (from Section~\ref{observe_data}) to ensure a realistic simulation. Then, we systematically test the single-component (\green{single \sersic}) and two-component (\green{double \sersic}) models using galaxy image fitting pipelines and testing their fitting performance.

The details of our simulation and fitting procedures are outlined in the rest of this section.
\subsection{\red{Generating and Modelling the Mock Galaxies}}\label{simu data}

\subsubsection{Parameter Settings}
Our simulations incorporate both bulge and disk components to represent the composite structure of real galaxies. To ensure our simulations encompass a wide range of realistic galaxy types and structural characteristics, we carefully set parameter ranges for the mock galaxies. These parameters include total magnitude, B/T, effective radius, axis ratios, and \sersic\ index, as summarized in Table~\ref{tab1}. Below, we provide a detailed justification for each parameter.

\textbf{Total Magnitude} ($m_{\text{total,sim}}$): \green{we set the total magnitudes of the mock galaxies to be uniformly distributed between 20 and 28 mag.} This range spans from moderately bright galaxies to faint galaxies near the detection limit of the CEERS survey~\citep{Finkelstein2022}. For context, a galaxy at redshift $z \approx 6.4$ with stellar mass $\log(M_*/M_\odot) \sim 11$ has observed magnitudes of $23.1$ and $25.1$ mag, in F356W and F150W respectively~\citep{2023Natur.621...51D}. 
By including faint galaxies (magnitude > 26), we account for the challenges of morphological decomposition at  low SNR, which are particularly {important} for high-redshift studies. This range ensures that our simulations are representative of the diverse luminosity distribution observed in both nearby and distant galaxy populations.

\textbf{Bulge-to-total ratios (B/T)}: \green{we set the B/T to be uniformly distributed between 10\% and 90\%.} While previous studies of nearby galaxies~\citep[e.g.][]{Simard_2011} typically report B/T ranges of 20\%--80\%\ for spiral galaxies, our broader distribution (10\%--90\%) intentionally encompasses the full theoretical range of bulge+disk configurations. This range covers a broad spectrum of galaxy types, from disk-dominated ($B/T< 0.5$) to bulge-dominated ($B/T > 0.5$) galaxies.
By including galaxies with comparable bulge and disk luminosities ($B/T \approx$ 50\%), we can assess the performance of decomposition techniques in cases where both components contribute significantly to the total light, which is the most distinctive cases against the scenario for single-component models.

\textbf{Bulge and Disk Effective Radius}:
We set the effective radius of the bulge component, $R_{\text{bulge,sim}}$, to follow a uniform distribution between 0\farcs05 and 0\farcs2 (corresponding to 0.4--1.6 kpc at $z\sim2$), consistent with the observed range of bulge sizes in both local galaxies~\citep[e.g.,][]{gadotti2009, fisher2008} and massive galaxies at high redshift~\citep[e.g.,][]{bruce2014, lebail2024}.
Compact bulges ($R_{\text{bulge}} < 0\farcs1$) are particularly relevant for high-redshift studies, where galaxies tend to be smaller and more compact. This range aligns with observations of compact bulges in high-redshift galaxies ($z>1$), where stellar components are typically confined to sub-kpc scales, e.g., $R_\text{eff}\sim0.5$ kpc in quiescent galaxies~\citep[e.g.,][]{vanderWel2014,Newman2012}. The disk effective radius ($R_{\text{disk,sim}}$) is set to 1.5--3.0 times the bulge radius, ensuring physically plausible proportions between the two components. This range captures the observed scaling relations between bulge and disk sizes in spiral galaxies, where disks are typically more extended than bulges -- a trend observed in both local spirals and $z \sim 1\text{--}3$ star-forming galaxies~\citep[e.g.][]{Sanchez2016,vanderWel2014}.

\textbf{Axis Ratios:}  
\green{we set the axis ratios for the bulge ($q_{\text{bulge, sim}}$) and disk ($q_{\text{disk, sim}}$) to be uniformly distributed.
For the bulge, we set $q_{\text{bulge, sim}}$ to range from 0.8 to 0.9, representing moderately flattened spheroidal shapes consistent with classical bulges in elliptical galaxies.
For the disk, we set $q_{\text{disk, sim}}$ to range from 0.4 to 0.7, capturing a diversity of moderate inclination angles commonly observed in spiral and irregular galaxies.} This selected range includes nearly face-on disks ($q \approx 0.7$) and moderately inclined systems ($q \approx 0.4$), thus avoiding both unrealistic completely face-on configurations ($q > 0.8$; observationally rare due to geometric selection effects, e.g., \citealt{10.1093/mnras/258.2.404}) and fully edge-on systems ($q \approx 0.1$ -- $0.2$; corresponding to intrinsic stellar disk thickness ratios $h_z/R_d \approx 0.1$ -- $0.2$, e.g., \citealt{van_der_Kruit_2011}). This ensures our simulations reflect realistic projection effects on morphological measurements without including the most extreme edge-on scenarios.

\textbf{\sersic\ index}:
The \sersic\ index for the bulge ($n_{\text{bulge,sim}}$) is fixed at 4.0, corresponding to a de Vaucouleurs profile~\citep[e.g.,][]{1948AnAp...11..247D}, which is typical for classical bulges in elliptical galaxies. For the disk ($n_{\text{disk,sim}}$ ), the \sersic\ index is fixed at 1.0, representing an exponential profile characteristic of \green{disky spiral galaxies}. 

By adopting this simplified framework, we aim to assess how well the true properties of galaxies can be recovered under the assumption that these idealized profiles accurately describe the underlying structure. This serves as a critical first step in understanding the performance of bulge+disk decomposition techniques in JWST observations, providing a baseline for future studies that may incorporate more complex structural components and evolutionary scenarios (see Section~\ref{first_step} for more discussions).

The detailed methodology for generating these mock galaxies is described in the following section.

\begin{table}
    \centering
    \caption{Parameters for simulated bulge+disk galaxies. All parameters are uniformly distributed within the given ranges; fixed parameters remain constant.}
    \label{tab1}
    \begin{tabular}{ll}
        \hline
        \textbf{Parameter} & \textbf{Range} \\
        \hline
        Total Magnitude ($m_{\text{total,sim}}$) & 20--28~mag \\
        B/T           & 10\%--90\% \\
        Bulge Effective Radius ($R_{\text{bulge,sim}}$) & 0\farcs05--0\farcs2 \\
        Disk Effective Radius ($R_{\text{disk,sim}}$)   & 1.5--3.0 times the bulge radius \\
        Axis Ratio for Bulge ($q_{\text{bulge,sim}}$)   & 0.8--0.9 \\
        Axis Ratio for disk ($q_{\text{disk,sim}}$)     & 0.4--0.7 \\
        Bulge \sersic\ Index ($n_{\text{bulge,sim}}$)   & Fixed (4.0) \\
        Disk \sersic\ Index ($n_{\text{disk,sim}}$)     & Fixed (1.0) \\
        \hline
    \end{tabular}
\end{table}

\subsubsection{generating noiseless image}
We adopt the image modelling \green{package} {\sc lenstronomy} to generate noiseless images of our simulated galaxies, ensuring that they match the pixel scale of the \green{CEERS mosaic} (i.e., 0$\farcs$03 per pixel). To account for convolution effects in our mock data, we incorporate a PSF model. For our simulations, we prepare five PSFs from the F356W band and four from the F150W band (see Section~\ref{observe_data}). In each realization, we randomly select one PSF to convolve the galaxy image, ensuring that the mock data accurately reflects realistic observational conditions,~\green{as illustrated in panel a of Figure~\ref{fig:1}}.

\subsubsection{Adding noise}
Then, we incorporate realistic observational noise conditions, including both Poisson noise and random background noise, into our noiseless image data. The Poisson noise is applied to the simulated images based on the exposure time of the CEERS observations 
and the gain values specific to JWST's NIRCam filters (2.0 e$^-/$ADU for SW, 1.8 e$^-/$ADU for LW).
These gain values, combined with the exposure time, determine the level of shot noise expected in the observed data, ensuring that our simulations accurately replicate the noise characteristics.

\begin{figure*}
\centering
\begin{tabular}{ccc}
\subfloat[High resolution bulge+disk image]{\includegraphics[trim = 0mm 0mm 0mm 0mm, clip,width=0.27\textwidth]{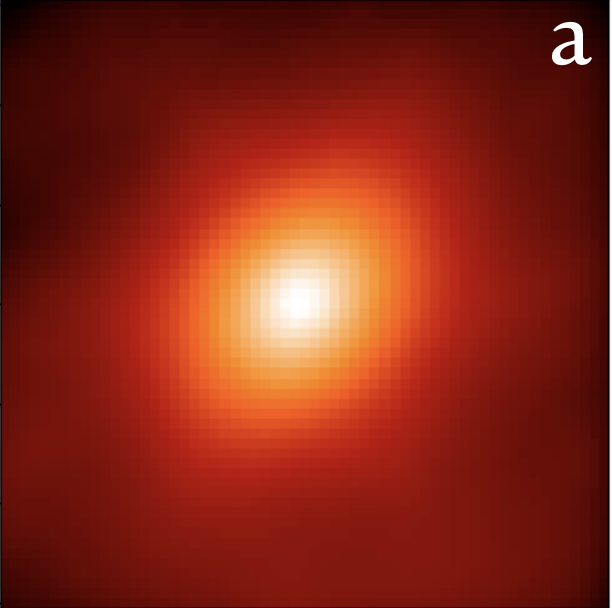}}&
\subfloat[Adding Poisson noise]{\includegraphics[trim = 0mm 0mm 0mm 0mm, clip,width=0.27\textwidth]{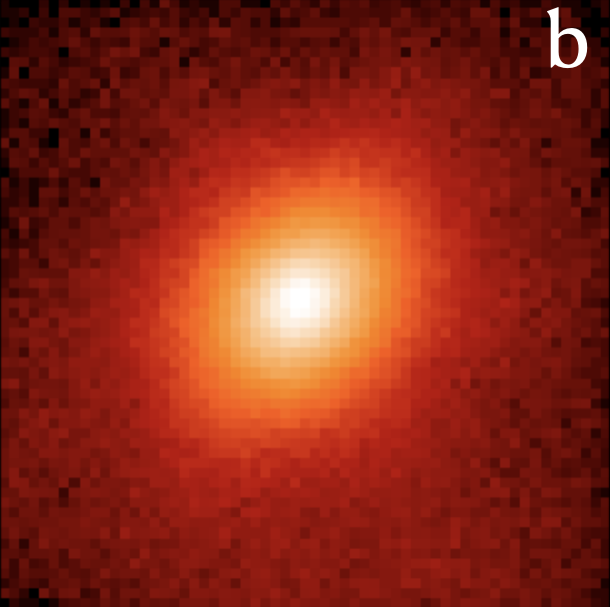}}&
\subfloat[Embedded into CEERS image]{\includegraphics[trim = 0mm 0mm 0mm 0mm, clip,width=0.27\textwidth]{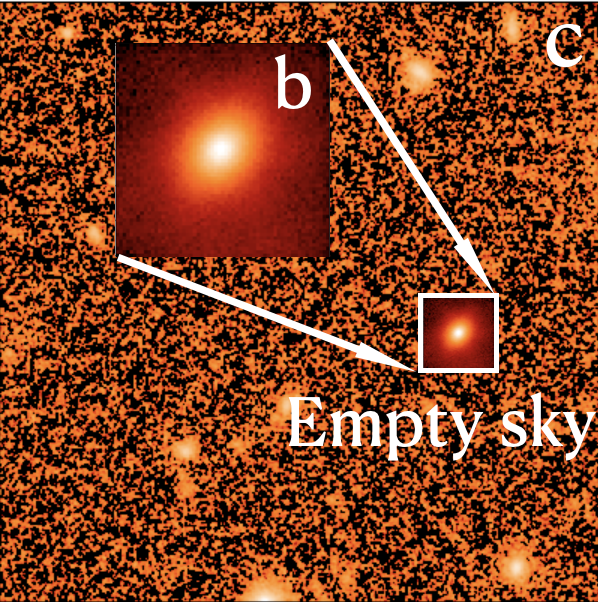}} 
\end{tabular}
\caption{This three-panel figure illustrates a systematic workflow for generating our mock galaxy images. Panel (a) shows a high-resolution, noise-free simulated galaxy, clearly displaying its structural morphology. Panel (b) introduces Poisson noise into the simulation, replicating realistic observational conditions encountered in astronomical imaging. Panel (c) presents a section of a CEERS image, into which we embed the simulated galaxy of panel (b) at a randomly selected empty sky location, indicated by the white square. This procedure allows comprehensive photometric measurements and accurate characterization of noise properties.}
\label{fig:1}
\end{figure*}

After accounting for the Poisson noise, we insert the mock image stamp into a randomly selected empty sky region among the CEERS field of view, as illustrated in Figure~\ref{fig:1}. This step allows the mock image to reflect the noise characteristics, background variations, and other observational factors encountered when analyzing real galaxies. In our subsequent fitting analysis, we will create new image cutouts at this position. The details of the fitting process are described in the following subsection.

\subsection{Fitting the mocks}\label{galight}
To analyze our mock galaxies, we employ two modeling approaches. First, we apply the {\it single \sersic\ model} that describes the galaxy's light profile as defined Equation \ref{sersic}. Second, we implement the {\it double \sersic\ model} that is composed of the bulge component (fixed at $n=4$ for classical bulges) and the disk component (fixed at $n=1$ for exponential disks); the two components will be modelled simultaneously during the fitting.

We use the \texttt{galight}~\citep[e.g.,][]{Ding_2020} package to perform the image fit. \texttt{galight} is a Python-based open-source package designed for two-dimensional model fitting of optical and near-infrared images. It characterizes the light distribution of galaxies by decomposing them into components such as disks, bulges, bars, and quasars. The package leverages the advanced image modelling capabilities of \texttt{lenstronomy} ~\citep[e.g.,][]{2018PDU....22..189B,2021JOSS....6.3283B}, with a redesigned user interface that facilitates automated fitting workflows. 

We use \texttt{galight} to prepare our modelling ingredients, which include the following: (1) image re-cutout with size six times to the effective radius of the simulated galaxy disk to capture low surface brightness features while avoiding too many empty pixels which could lower-weight the SNR while bias the fitting, (2) a noise level map describing the uncertainty of each pixel, combining Poisson noise (estimated from the exposure time and gain values) and background noise (measured from nearby blank sky regions), and (3) a PSF from our PSF collection.

It is worth noting that, during the fitting stage, we account for PSF uncertainties by implementing a mismatch strategy. Specifically, we simulate the galaxy image using one PSF but apply a different PSF during the fitting process. This approach allows us to systematically evaluate the sensitivity of our results to PSF variations and identify potential systematic errors introduced by PSF mismatches. By quantifying the impact of PSF uncertainties, we gain a deeper understanding of their effects on photometric, morphological, and decomposition analyses, ensuring robust and reliable conclusions.

During the fitting, we involve two modelling configurations: the single \sersic\ model and the double \sersic\ models. For the single \sersic\ models fitting, the index value is limited to the range of 1--4. For the double \sersic\ models fitting, the \sersic\ indices $n$ of the bulge and disk components are fixed to 4 and 1, respectively, \green{ while all other parameters remain free to vary during optimization.} To help the fitting parameter converge quickly, we use the truth parameters that were used when generate the mock galaxy as the initial inputs during the fitting.\footnote{We also tested to use the default parameters as introduced by \texttt{galight} based on the initial morphology of the fitting targets as the initial input. We find that both initial inputs yield consistent results.} For the parameter minimization, we employ the Particle Swarm Optimization (PSO) algorithm~\citep[e.g.,][]{Kennedy1995}, consisting of two shallow runs followed by one deep run, to ensure convergence.

In addition to using \galight, we also test the performance of the popular galaxy image modelling software \texttt{galfit} ~\citep[e.g.,][]{peng2002detailed} to fit the mock images. This comparison will allow us to determine whether the two independent codes yield consistent results. By evaluating the outputs from both software packages, we can assess their reliability and robustness in modelling galaxy photometry and morphology (comparison results are presented in Section~\ref{compare_galfit}).

\subsection{Robustness of Simulation and Fitting Tests}
To ensure the statistical robustness of our conclusions, we generated a dataset of 5,000 simulated galaxies, as introduced in Section~\ref{simu data}. This sample size provides comprehensive coverage of the parameter space as defined, allowing us to reliably assess the performance of single \sersic\ and double \sersic\ models decomposition models across a wide range of galaxy morphologies and observational conditions.

We assess the accuracy of the recovered galaxy properties, such as magnitudes and effective radius, by comparing them to their intrinsic simulated values. This enables us to systematically identify trends and quantify biases in these key properties across a range of galaxy conditions.

\section{Results}\label{sec.4}

\begin{figure*}
    \centering
    \begin{subfigure}{1.05\textwidth}
        \centering
        \includegraphics[width=\textwidth]{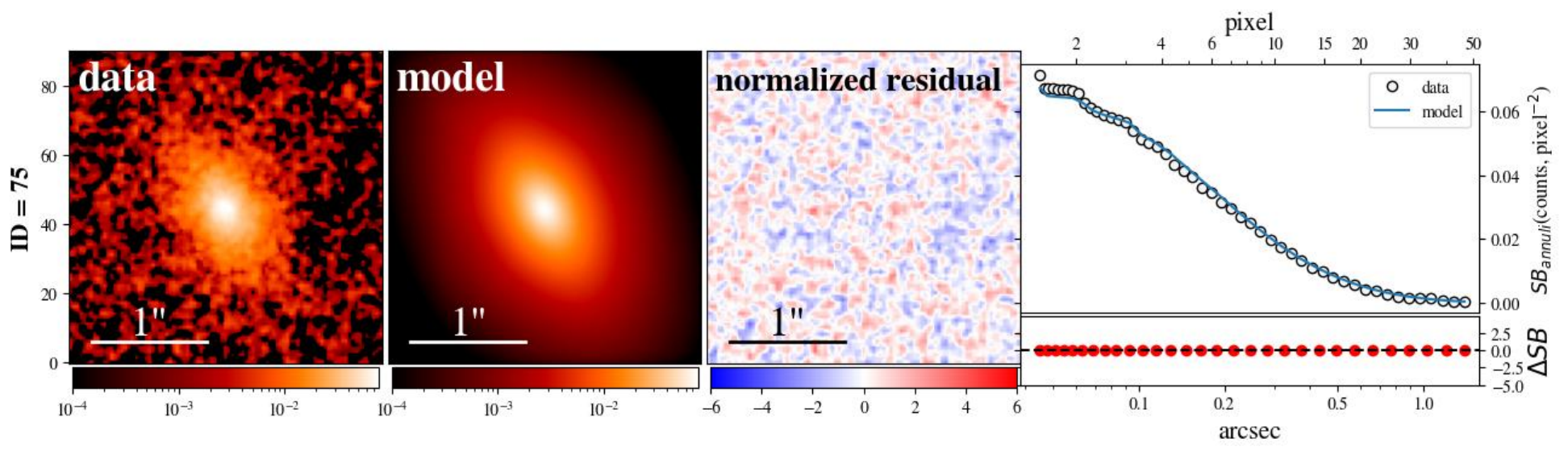}
        
        \label{fig:first_image}
    \end{subfigure}
   \vskip -1.75em
    \begin{subfigure}{1.05\textwidth}
        \centering
        \includegraphics[width=\textwidth]{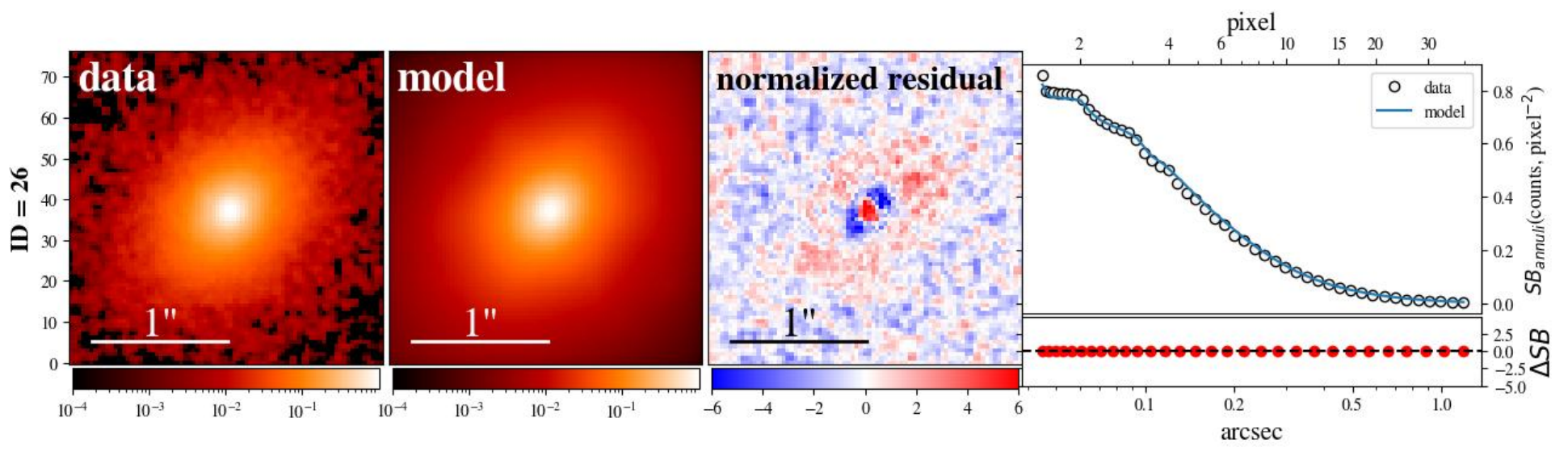}
        \label{fig:second_image}
    \end{subfigure}
    
   \caption{Single \sersic\ model fitting results for two example simulated galaxies with B/T of 20\% (ID = 75) and 50\% (ID = 26), respectively. The panels from left to right are: (1) the simulated galaxy image, (2) the best-fitting Single \sersic\ model, (3) the residuals normalized by the variance, and (4) the one-dimensional surface brightness profiles (top) with the corresponding residuals (bottom). In panel (4), the open circles represent the simulated data, while the blue line corresponds to the best-fitting model. Note that the one-dimensional profiles are for illustration only, as the actual fitting process is performed using the two-dimensional images.}
   \label{fig:2}
\end{figure*}

\subsection{Single \sersic\ Model Fitting Results}\label{single_result}
We begin by evaluating single \sersic\ fits to our bulge+disk mock galaxies with three key goals: (1) to determine whether single-component models can reveal the presence of two-component systems through diagnostic features (e.g., residual maps). (2) to quantify how accurately single \sersic\ parameters -- such as the total magnitude, effective radius,  (3) and explore the correlation between \sersic\ index and B/T ratio.

\begin{figure*}
    \centering
    \includegraphics[width=1.05\textwidth]{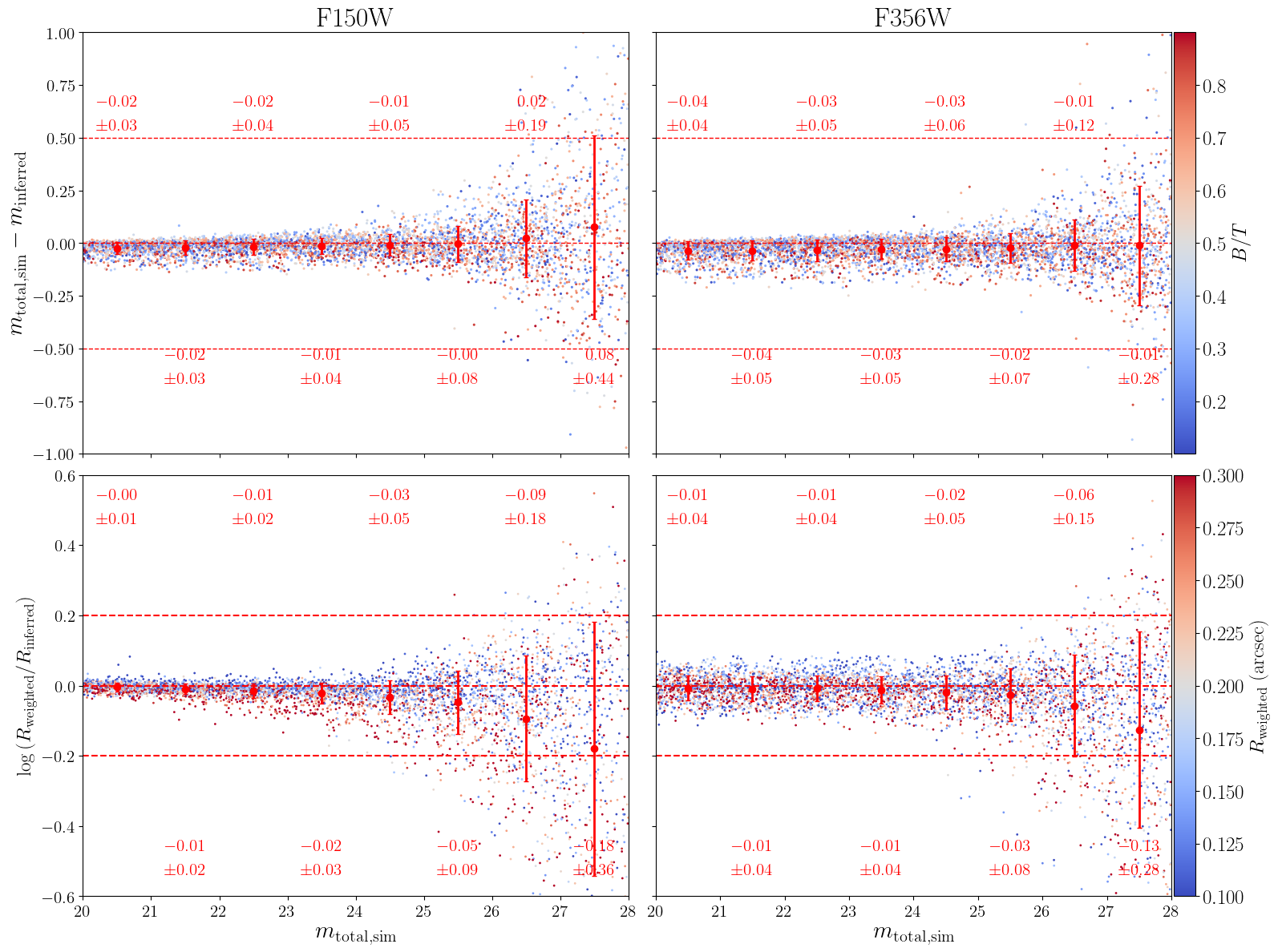}
    \caption{Parameter recovery results for single \sersic\ fits. The left column shows the results for the F150W band, while the right column corresponds to the F356W band. The color bars represent the B/T  (top panels) and the simulated galaxy radius $R_{\text{sim}}$ (bottom panels, as defined in Equation~\ref{eq:2}). For simulated galaxies brighter than magnitude 26 ($m_{\text{total,sim}}<26$), fitting accuracies are high, with magnitude errors consistently within 0.1 mag and size errors within 0.1 dex for both bands. From the magnitude comparisons, we observe no clear correlation between B/T and magnitude errors, suggesting that magnitude recovery is independent of the bulge-to-total ratio. To enhance visual clarity, the color scale for the size comparison plots (second row) is limited to a range of 0\farcs1–0\farcs3; note that this range does not represent the actual minimum and maximum sizes but is chosen purely to emphasize the correlation between size and recovery accuracy by mapping the smallest and largest simulated sizes to 0\farcs1 and 0\farcs3, respectively. Galaxies with larger radii (redder points) generally exhibit more accurate size recovery.}
    \label{fig:3}
\end{figure*}

\subsubsection{Presence of hidden structure in residual map }
We first examine a fitting result of a galaxy (ID = 75) with a B/T of approximately 20\%, indicating that the system is dominated by one structural component, as shown in Figure~\ref{fig:2}  (top panel). In this scenario, the light profile of a single \sersic\ model reproduces this galaxy, showing negligible residuals and demonstrating its effectiveness for single-component systems. In contrast, for a galaxy with a B/T of roughly 50\% (Figure~\ref{fig:2} bottom panel, ID = 26), where bulge and disk contributions are comparable, the single \sersic\ fit shows significant residuals -- particularly in the central region.

\red{These results underscore two important insights: first, a good fit with a single \sersic\ model, indicated by small residuals, does not necessarily imply a genuinely single-component galaxy -- it could instead reflect a two-component structure with a dominant component masking the presence of a subdominant one. Second, noticeable residuals in single \sersic\ fits strongly suggest the presence of additional structural complexity, reinforcing the necessity and value of employing double \sersic\ models (such as bulge+disk) models for accurate galaxy decomposition.}

\begin{figure*}
    \centering
    \includegraphics[width=\textwidth]{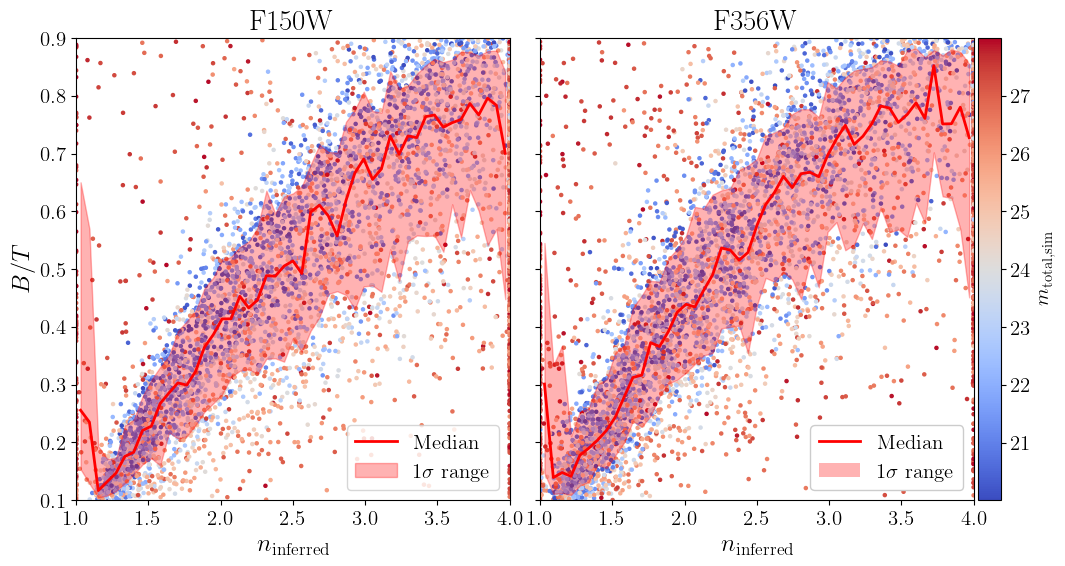}
    \caption{Relation between \sersic\ index (fitting) and simulated B/T (truth) for galaxies in the F150W and F356W. Scatter points represent individual galaxies, with colors indicating their total magnitudes. The red line represents the median B/T value at each \sersic\ index, and the shaded region shows the 68.27\% uncertainty interval. The consistent trends in F150W and F356W suggest that the correlation between $n$ and B/T is universal across different bands.}
    \label{fig:8}
\end{figure*}

\subsubsection{Evaluating magnitude and size measurements}
We now estimate the performance of how single \sersic\ fitting results capture the overall properties of the bulge+disk system by making a comparison of the inferred galaxy properties with the true input values from our simulations. 
Figure~\ref{fig:3} (top) shows the comparison of total magnitudes between the simulated  values (input) and those recovered using the single model \sersic\ (inferred) for the F150W and F356W filters. The mean and standard deviation of the magnitude offsets are indicated for each magnitude bin. For galaxies brighter than 26 mag, the biases and scatters are small, with offsets distributed as \( -0.015 \pm 0.049 \)~mag in F150W and  \( -0.031 \pm 0.053 \)~mag in F356W, demonstrating that the single \sersic\ model can accurately recover the overall truth magnitudes. However, for fainter galaxies (mag $> 26$), the scatter increases, with mean residuals of \( 0.047 \pm 0.323 \)~mag in F150W and \( -0.011 \pm 0.219\)~mag in F356W. This increased scatter reflects the challenges of fitting faint sources, where lower SNR leads to greater uncertainties in parameter recovery. Despite the degradation for fainter galaxies, the overall residuals remain far below 0.5 magnitudes, even for galaxies as faint as 28 mag.

Intriguingly, we observe no strong correlation between the B/T and magnitude residuals, indicating that the single \sersic\ model effectively recovers the total brightness of galaxies regardless of their structural complexity. While the model struggles to resolve individual bulge and disk components in systems with comparable B/T  (e.g., the bottom panel of Figure~\ref{fig:2}), its ability to capture the integrated light remains robust.

We also compare the fitting results for galaxy sizes, specifically the effective radius (\reff), between the simulated values and those obtained from the fitting process, as shown in Figure~\ref{fig:3} (bottom). Since our simulations use bulge+disk components, there is no single effective radius for direct comparison with the single \sersic\ fitting. Instead, we calculate a flux-weighted effective radius for the simulated galaxy properties, denoted as \(R_\text{weighted}\), using:
\begin{equation}
    \label{eq:2}
R_\text{weighted} = \dfrac{R_\text{bulge,sim}  f_\text{bulge,sim} + R_\text{disk,sim}  f_\text{disk,sim}}{f_\text{bulge,sim} + f_\text{disk,sim}},
\end{equation}
where \(f_\text{bulge,sim}\) and \(f_\text{disk,sim}\) represent the fluxes of the simulated bulge and disk, respectively.

In Figure~\ref{fig:3} (bottom), for bright galaxies (mag $< 26$), the flux-weighted radius agree well with the \sersic\ radius, with residuals of \( -0.021 \pm 0.046 \)~dex in F150W and \( -0.013 \pm 0.047 \)~dex in F356W, indicating accurate measurements in high-SNR conditions. For faint galaxies (mag \(> 26\)), the \green{offsets} increase significantly, reaching \(-0.132 \pm 0.280\)~dex in F150W and \(-0.092 \pm 0.226\)~dex in F356W. While the fitting accuracy decreases at low SNR, we observe an interesting trend: the F150W band, despite its higher spatial resolution, exhibits comparable or even larger deviations in the inferred \( R_{\text{eff}} \) compared to F356W.

This behavior can be explained by the interplay of spatial resolution, PSF effects, and background noise. 
For bright galaxies (\( \text{mag} < 26 \)), the higher resolution in F150W allows more precise recovery of galaxy structure, resulting in lower residuals. The higher spatial resolution of F150W is particularly advantageous for resolving compact structures like bulges in early-type galaxies, which improves the accuracy of \sersic\ profile fits.
\green{For faint galaxies ($\text{mag} > 26$), PSF mismatches and noise propagation become significant. The lower sensitivity in F150W leads to larger uncertainties in inferred sizes at lower flux levels. In contrast, the PSF at longer wavelengths (F356W) is generally more stable because optical aberrations, alignment errors, and scattering effects are less pronounced, making the PSF shape more predictable and consistent. Consequently, F356W provides more precise parameter recovery for faint galaxies.}

\subsubsection{Single \sersic\ index ($n$) as a proxy for B/T}
\label{dis:n}

We further examine the distribution and implications of the inferred \sersic\ index, analyzing how these values correlate with B/T values for galaxies with bulge-to-disk structure, as shown in Figure~\ref{fig:8}. The \sersic\ index ($n$) has long served as a convenient estimator for galaxy morphology, with $n\approx1$ signaling disk-dominated systems and $n\approx4$ indicating bulge-dominated ellipticals~\citep[e.g.,][]{de1948recherches, 1970ApJ...160..811F,2009ApJS..182..216K}. Observational studies of high-redshift galaxies sometimes adopt $n$ as a proxy for B/T ratios when double \sersic\ decomposition is infeasible due to resolution or SNR limitations~\citep[e.g.,][]{Ding_2020}. This approach implicitly assumes a monotonic relationship between $n$ and B/T, where higher $n$ values correlate with more bulge-dominated systems -- a premise grounded in local galaxy studies.

However, this simplification carries risks: single-component \sersic\ fits conflate structural complexity (e.g., bulge+disk systems) into a single parameter, potentially biasing interpretations of galaxy evolution. Our simulations -- with known bulge and disk properties -- 
 provide the first controlled test of this assumption under JWST-like conditions.

We collected the \sersic\ $n$ values obtained from the single \sersic\ fitting and compared them with the truth value of the B/T in the system. 
The results, as shown in Figure~\ref{fig:8}, reveal a strong correlation between the fitted $n$ and the true B/T, with consistent trends across the F150W and F356W bands. The overlapping median lines and 1$\sigma$ ranges in both bands demonstrate the robustness of this relationship. As expected, the single \sersic\ index serves as a reliable proxy for the B/T  in our simulations. When the true B/T  increases from 10\% to 90\%, the fitted $n$ values systematically rise from 1.0 to 4.0, reflecting the transition from disk-dominated to bulge-dominated systems. However, when the fitted $n$ approaches its upper or lower limits (1 or 4), the scatter in the B/T  increases significantly, as the model becomes less sensitive to subtle variations in galaxy structure. Despite this, the consistent trends in Figure~\ref{fig:8} (F150W and F356W) suggest that the correlation between $n$ and B/T is universal across different bands.

We also identify several notable limitations to this correlation. First, we observe that when the total host magnitude is brighter than 24.5, the correlation becomes tighter, indicating that a sufficient SNR  is crucial for maintaining the validity of this relationship. In contrast, for fainter hosts (i.e., those with magnitudes greater than 26), the scatter in the data increases significantly. Additionally, our simulations assume that the bulge has a \sersic\ index ($n$) of 4, meaning that this proxy is applicable only to classical bulges and does not hold for the pseudo-bulges. Furthermore, external factors such as dust and projection effects can introduce biases in the measured parameters, particularly at shorter wavelengths and higher inclinations. These limitations highlight the need for caution when interpreting the \sersic\ index as a proxy for the B/T, emphasizing the importance of considering other morphological indicators alongside it to achieve a more comprehensive understanding of galaxy structure.

\begin{figure*}
    \centering
        \includegraphics[width=1.05\textwidth]{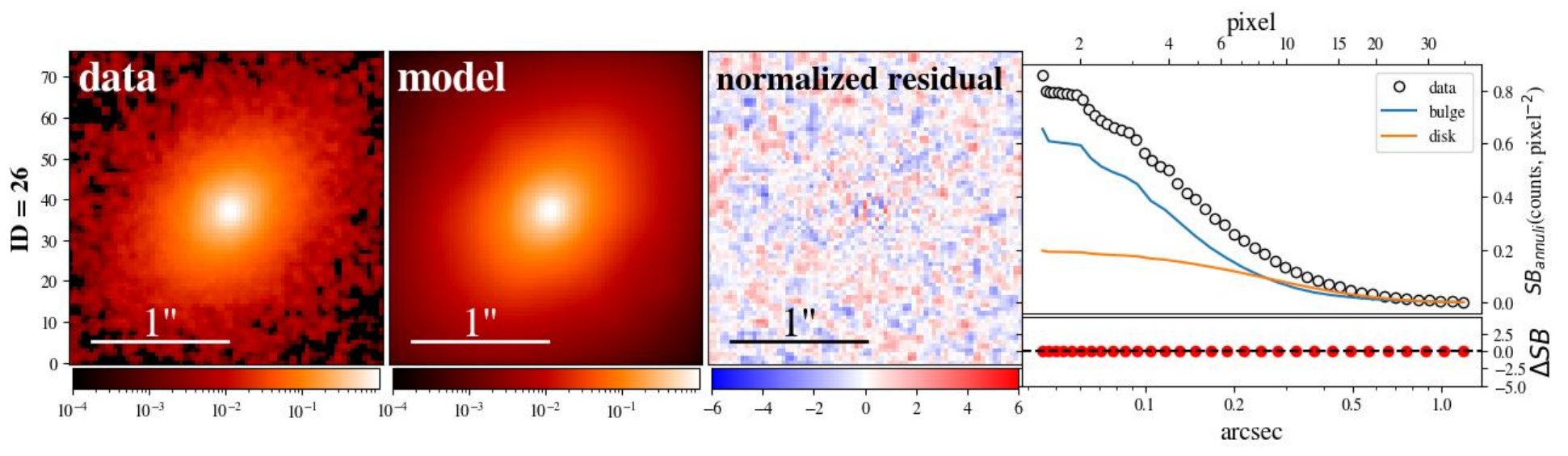} 
       \caption{Double \sersic\ model fitting results for the simulated galaxy ID = 26. The blue and orange lines in panel (4) represent the bulge and disk contributions, respectively, while the black line represents the total model profile. See the caption of Figure~\ref{fig:2} for details.}
        \label{fig:4}
\end{figure*}

\subsection{Double \sersic\ models Fitting Results}
We now consider the decomposition scenario in our fitting process, where we already know the galaxy is composed of  bulge and disk components. Our goal is to perform a double component decomposition, trying to accurately recover these distinct features within the galaxy. For example, in Figure~\ref{fig:4}, we present the double-component fitting results for the same galaxy (ID=26) as previously fitted based on a single \sersic\ in Figure~\ref{fig:2} (bottom). The results, as demonstrated by the residual map, indicate a significant improvement in the fit after applying the double \sersic\ model, demonstrating the advantages of accurately modelling both components in our analysis.

\subsubsection{Recovery of magnitude as a function of B/T}
We evaluate the precision and accuracy of parameter recovery by comparing the fitted bulge and disk magnitudes (inferred) with their true input values (sim) in Figure~\ref{fig:5} (top) for the F150W and F356W, respectively. Our results indicate that unbiased magnitude estimates are achievable for both the bulge and disk components across these two bands. Additionally, we observe an increase in uncertainties as the components become fainter, a trend also seen in the single \sersic\ scenario.  We find that the inferred magnitude uncertainty remains below 0.5 mag when the bulge or disk is brighter than 26 mag, suggesting that the decomposition fitting method remains robust for relatively bright galaxies. For brighter galaxies (mag < 26), the mean and standard deviation of the bulge magnitudes  are \( -0.048 \pm 0.194 \)~mag in F150W and \( -0.090 \pm 0.328 \)~mag in F356W. For fainter bulges (mag > 26 ), the scatter increases, with the mean residuals reaching \( -0.004 \pm 0.545 \)~mag in F150W and \( -0.023 \pm 0.507 \)~mag in F356W. 

Furthermore, when considering  the sample with B/T > 60\%, the values improve to \( -0.042 \pm 0.151 \)~mag in F150W and \( -0.082 \pm 0.203 \)~mag in F356W for brighter galaxies (mag $<26$). This phenomenon suggests that galaxies with a high B/T ($> 60\%$) are more likely to yield accurate estimates in luminosity fitting. In these galaxies, the bulge component is more prominent and contributes significantly to the overall luminosity, making it easier to obtain accurate bulge fits that are less influenced by other factors, such as the disk component or outer regions of the galaxy. These findings highlight the importance of considering the different systematics between various galaxy types when performing luminosity decompositions. Particularly in the case of faint or structurally complex galaxies. This trend is also apparent in the disk fitting: for galaxies with a B/T ratio of less than 0.4 and a magnitude of brighter than 26, the accuracy of the recovered disk magnitudes improves. Specifically, for the F150W band, the values change from \( -0.053 \pm 0.233\)~mag to \( -0.048 \pm 0.124\)~mag, and for the F356W band, they improve from \( -0.086 \pm 0.411\)~mag to \( -0.071 \pm 0.203\)~mag.

\subsubsection{Evaluation of size measurements}

\begin{figure*}
    \centering
    
     \includegraphics[width=1.05\textwidth]{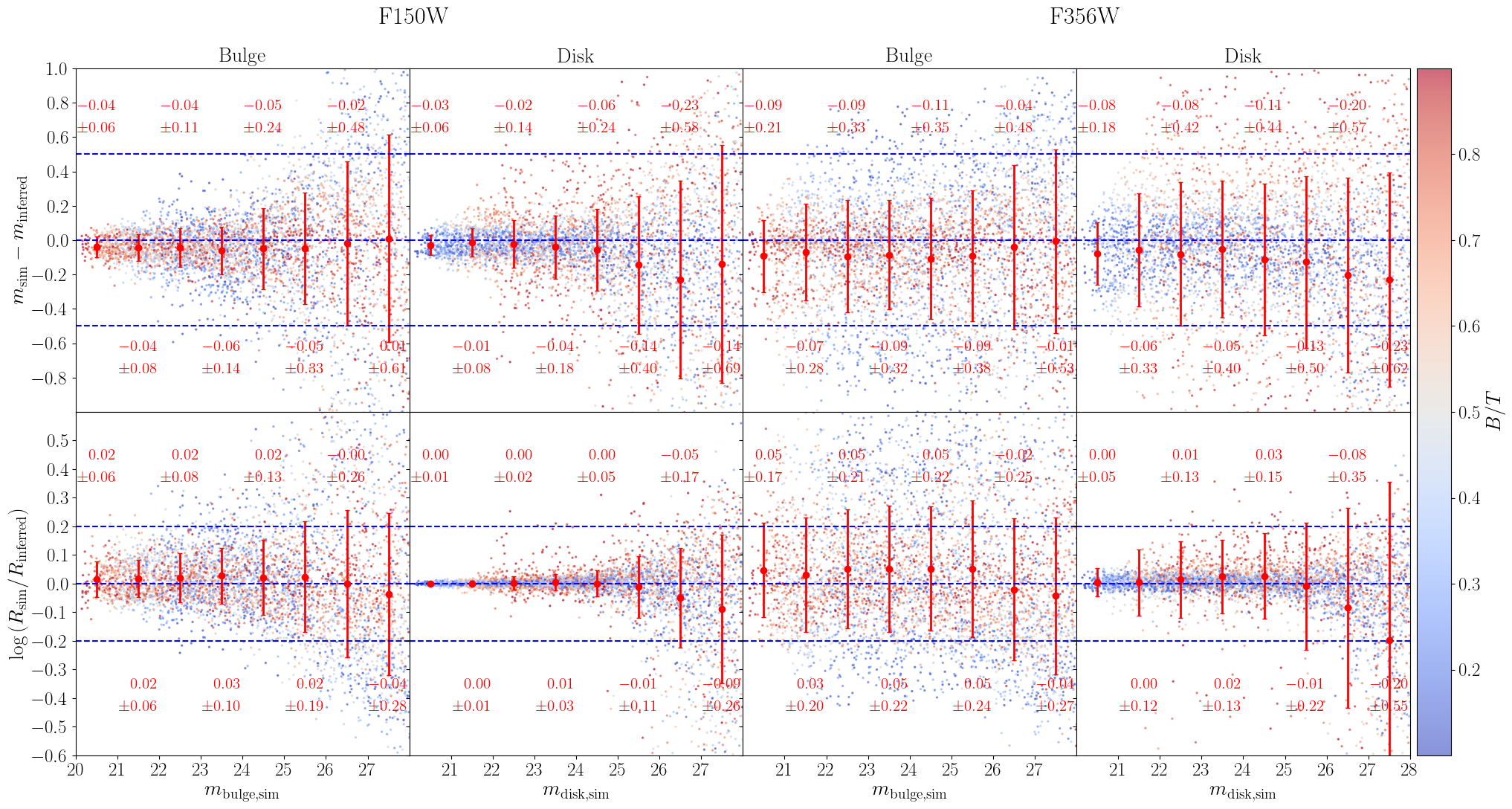}
      \caption{Comparison of simulated and inferred galaxy parameters: differences in magnitude and effective radius as functions of simulated bulge and disk magnitudes in the F150W and F356W filters. The color scale represents the B/T for both bulge and disk panels. The red points with error bars indicate the binned mean residuals and associated standard deviations, highlighting the reliability of the fitting procedure. These statistical values are annotated at relevant positions within the plots for clarity.}
    \label{fig:5}
\end{figure*}

The comparison of \reff\  is presented in the bottom part of Figure~\ref{fig:5}. Similarly, we find unbiased results for the global distribution for both the bulge and disk components. For the bulge's \reff, we observe uncertainties ranging from 0.17 to 0.24 dex for brighter galaxies in the F356W band, which are higher than the uncertainties seen in the F150W band, where they range from 0.06 to 0.19 dex. Again, this difference is likely due to the superior spatial resolution of the F150W filter, which yields more precise measurements of the bulge's effective radius. For both bands, we find that the systematic errors in the effective radius for the disk are generally smaller than those for the bulge. This observation can be attributed to the more extended nature of the disk component: the bulge, being more compact and centrally concentrated, is subject to greater variability in its measurement, especially in the presence of noise or in regions with less defined structures, leading to larger uncertainties. In contrast, the disk, with its larger and smoother profile, tends to yield more reliable estimates for its effective radius, thus resulting in smaller uncertainties.

In summary, our results demonstrate that bulge+disk decomposition techniques robustly recover the structural properties of both components. The precision of bulge and disk parameter recovery -- particularly in F150W, where angular resolution minimizes degeneracies between compact and extended structures -- validates the use of double \sersic\  modelling for JWST data. While single \sersic\ models remain effective for measuring total flux (Section~\ref{single_result}), our findings demonstrate that double \sersic\ models decomposition is essential for resolving the distinct physical processes shaping galaxies. This capability is critical for probing the evolutionary pathways of galaxies, as accurate bulge/disk separation enables direct tests of hierarchical assembly (e.g., bulge growth via mergers vs. disk growth via accretion).

\section{Discussion}\label{sec.5}

\subsection{Model Selection Using BIC: Single \sersic\ vs. Double \sersic\ models}
Having quantified the accuracy of parameter recovery for both single \sersic\ and double \sersic\ models, we now address a critical question: can we reliably identify which galaxies require a two-component decomposition? To determine this, we compare the relative performance of the single \sersic\ and double \sersic\ models models using the BIC. As a general principle, a lower BIC value indicates a better fit to the data, penalizing model complexity.

To quantify the improvement gained by using a double \sersic\  models, we define \( \Delta \text{BIC} = \text{BIC}_{\text{double \sersic }} - \text{BIC}_{\text{single \sersic}} \). A negative $\Delta \text{BIC}$ indicates that the double \sersic\ models provides a better fit. However, it's crucial to acknowledge that the reliability of BIC-based model selection depends on the SNR of the data. We therefore expect BIC to be a more robust indicator for brighter galaxies.

Indeed, we find a clear trend with magnitude. For galaxies with total light brighter than mag$ < 24.9$ in F150W and mag $< 26.2$ in F356W, 90\% of the targets yield $\Delta \text{BIC} < -10 $. These ranges change to mag $< 25.4$ (F150W) and mag $< 26.4$ (F356W), when 68\% of the targets yield  $\Delta \text{BIC} < -10 $.
This threshold suggests that these galaxies can be confidently classified as double \sersic\  systems based on the BIC criterion.
Furthermore, when the bulge and disk contribute comparable flux (30\% < B/T < 70\%), 90\% of the galaxies are correctly identified as double \sersic\ systems based on $\Delta \text{BIC} < -10 $.

Conversely, for fainter galaxies, or in cases where one component dominates the light profile (e.g., ID 75, Figure~\ref{fig:2}, top), the BIC values for the single \sersic\ and double \sersic\ models are comparable. This makes it challenging to confidently classify these galaxies as double \sersic\ models systems using BIC alone.

\subsection{Systematics as a function of SNR}
One of our goals is to provide guidance on the error budget for JWST-like conditions and offer a robust framework for quantifying decomposition systematics in the NIRCam surveys. To this end, we analyze the biases in recovered galaxy properties as a function of SNR, ensuring that our results are broadly applicable across different observational conditions. By systematically quantifying these biases, we aim to establish a reference that can be used not only for CEERs but also for other surveys with varying sensitivities.

The total SNR value of one galaxy is calculated as follows:
\begin{equation}
    \mathrm{SNR} = \frac{\sum \mathrm{flux}_i}{\sqrt{\sum \mathrm{noise}_i^2}}
    \label{eq:snr}
\end{equation}
where $\mathrm{flux}_i$ represents the pixelated flux distribution of astronomical sources, and $\mathrm{noise}_i$ denotes the corresponding noise components. This formulation explicitly accounts for the cumulative signal detection and noise propagation characteristics inherent in broadband photometric measurements.
\begin{figure*}
    \centering
    \includegraphics[width=1.05\textwidth]{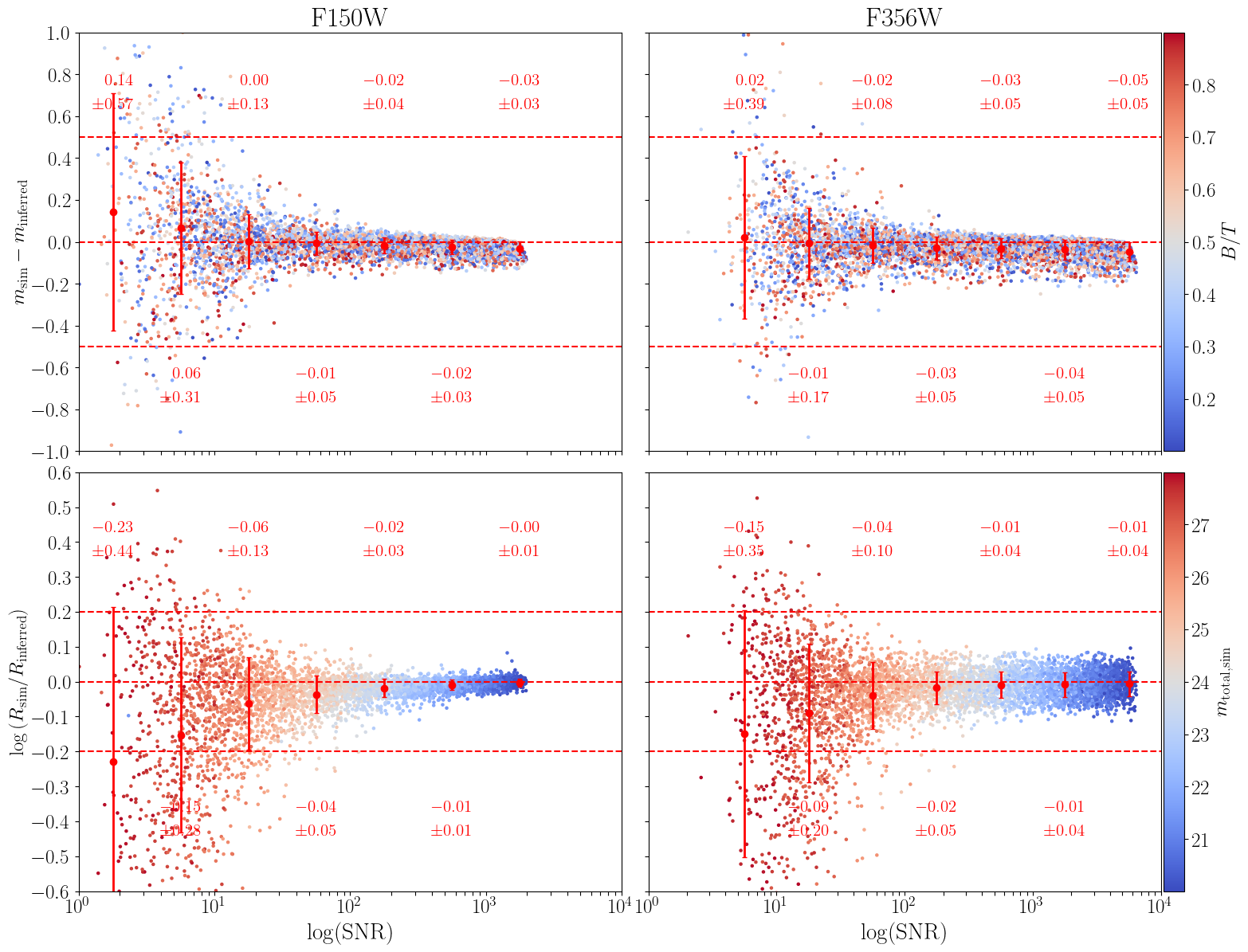}
    \caption{Comparison of simulated and inferred parameters for single \sersic\ fitting as a function of SNR. See the caption of Figure~\ref{fig:3}  for details.}
    \label{fig:6}
\end{figure*}
\begin{figure*}
    \centering
    \includegraphics[width=1.05\textwidth]{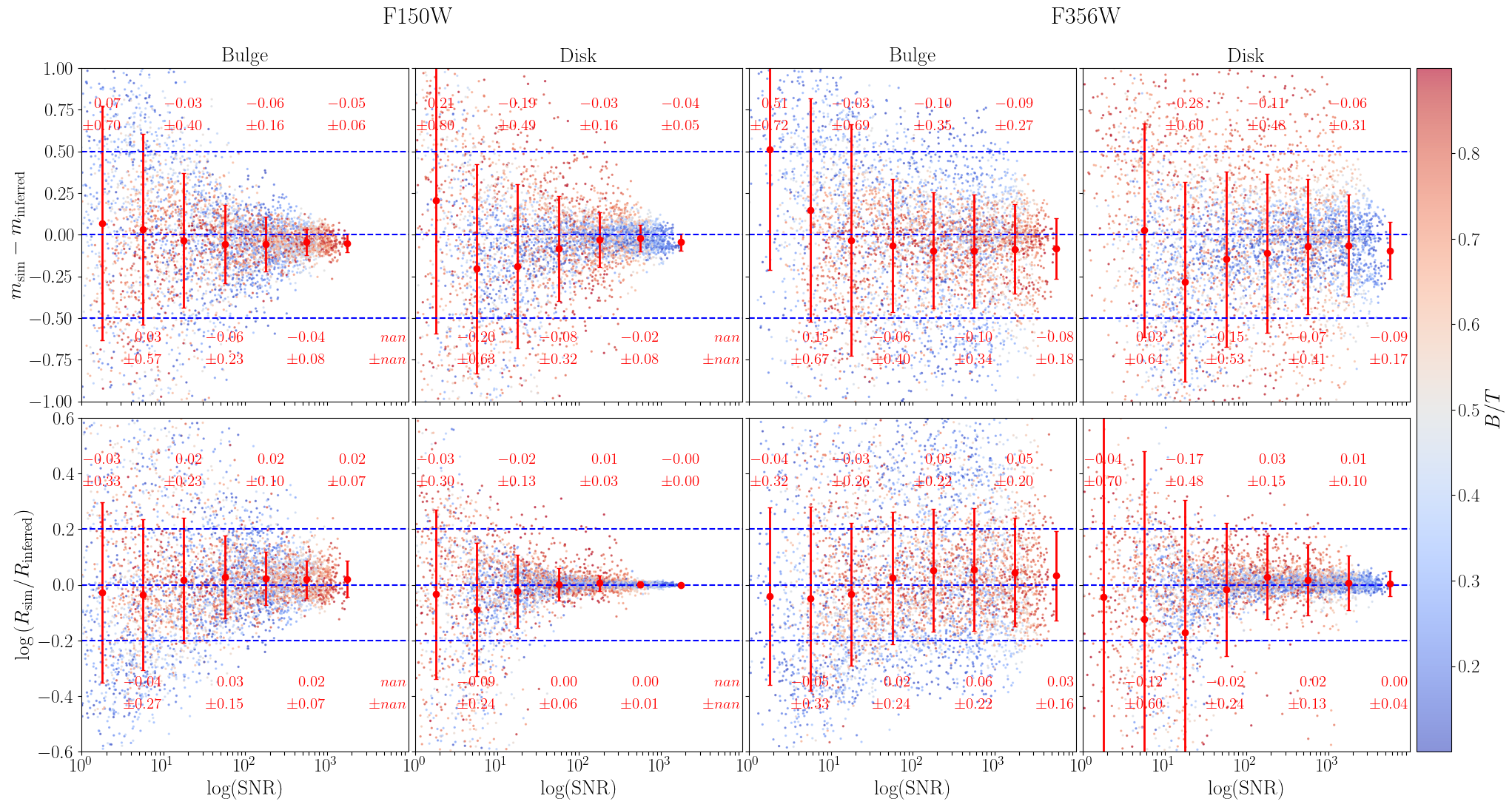}
    \caption{Comparison of simulated and inferred parameters for galaxy components as a function of SNR. See the caption of Figure~\ref{fig:5} for details.}
    \label{fig:7}
\end{figure*}

As illustrated in Figure~\ref{fig:6}, the single \sersic\ fitting results indicate that the scatters between fitted and simulated values for magnitude and effective radius decrease steadily with increasing SNR. This behavior aligns with our expectations: a higher SNR enables a more accurate reconstruction of the corresponding parameters. Furthermore, our analysis reveals that the accuracy of magnitude reconstruction is independent of the B/T.

For the double \sersic\ models fitting results shown in Figure~\ref{fig:7}, we observe a similar trend to the single \sersic\ model: the reconstruction accuracy of the corresponding parameters improves as the SNR increases. Likewise, the reconstruction accuracy for the bulge+disk decomposition is lower than that of the single \sersic\ fit. This is due to the increased complexity and inherent difficulty in accurately reconstructing parameters in a two-component model. Also, the results indicate that for the two-component decomposition, when the B/T is large (i.e., bulge flux dominates), the bulge parameters are recovered more accurately. Conversely, when the B/T is small (i.e., disk flux dominates), the disk parameters exhibit higher reconstruction accuracy.

For reference, we provide detailed numerical results in the corresponding figures.
By providing these error budgets for JWST-like conditions (using F150W and F356W as representative short- and long-wavelength filters), these results offer a practical framework for evaluating systematics in other NIRCam photometric studies.

\subsection{Comparison with Galfit} \label{compare_galfit}
\begin{figure*}
    \centering
    \includegraphics[width=1.05\textwidth]{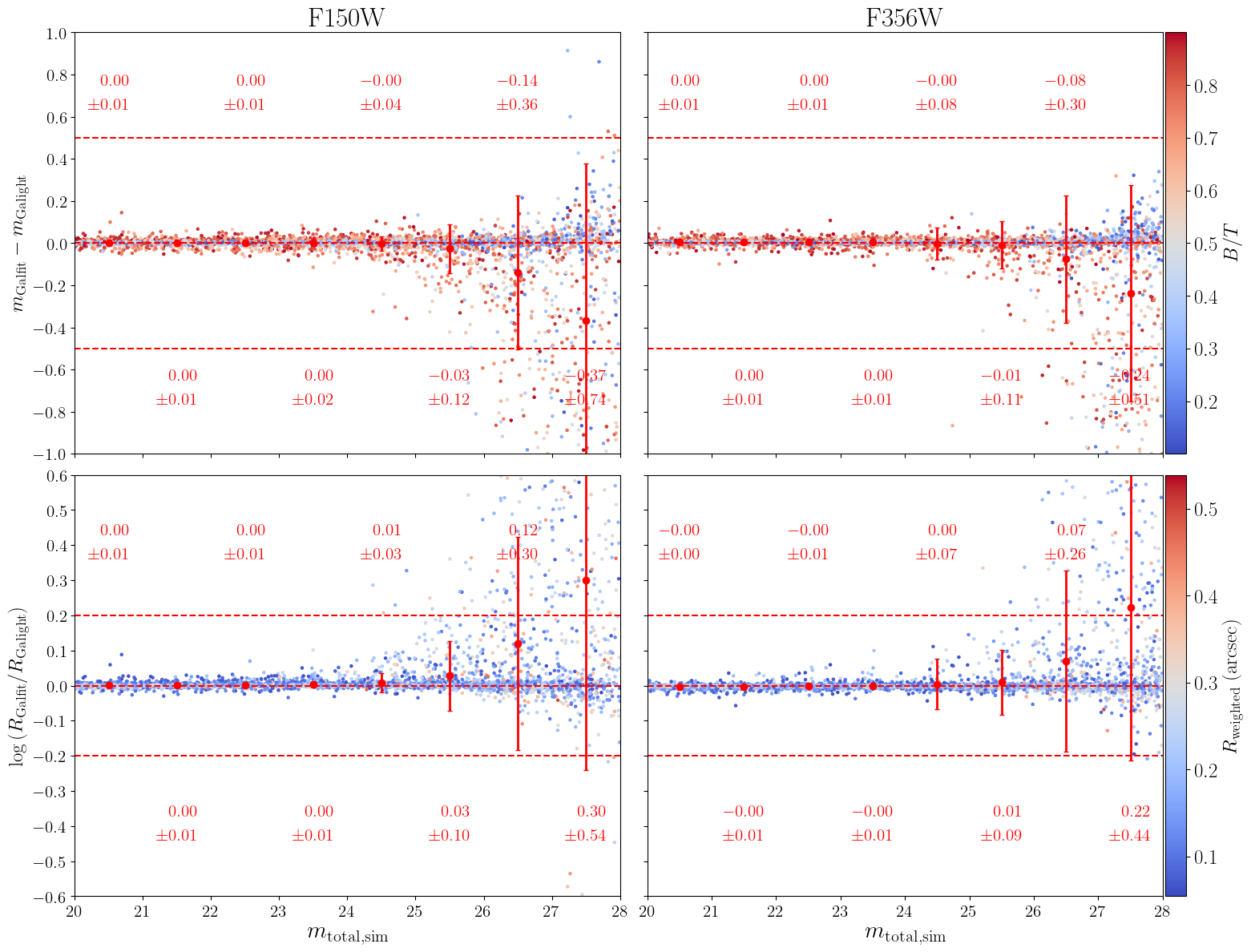}
 
    \caption{Comparison between \texttt{galfit} and \galight\ for magnitude and effective radius of single \sersic\  model fitting . See captions of Figure~\ref{fig:3}  for details.}
    \label{fig:9}
\end{figure*}

\begin{figure*}
    \centering
    \includegraphics[width=1.05\textwidth]{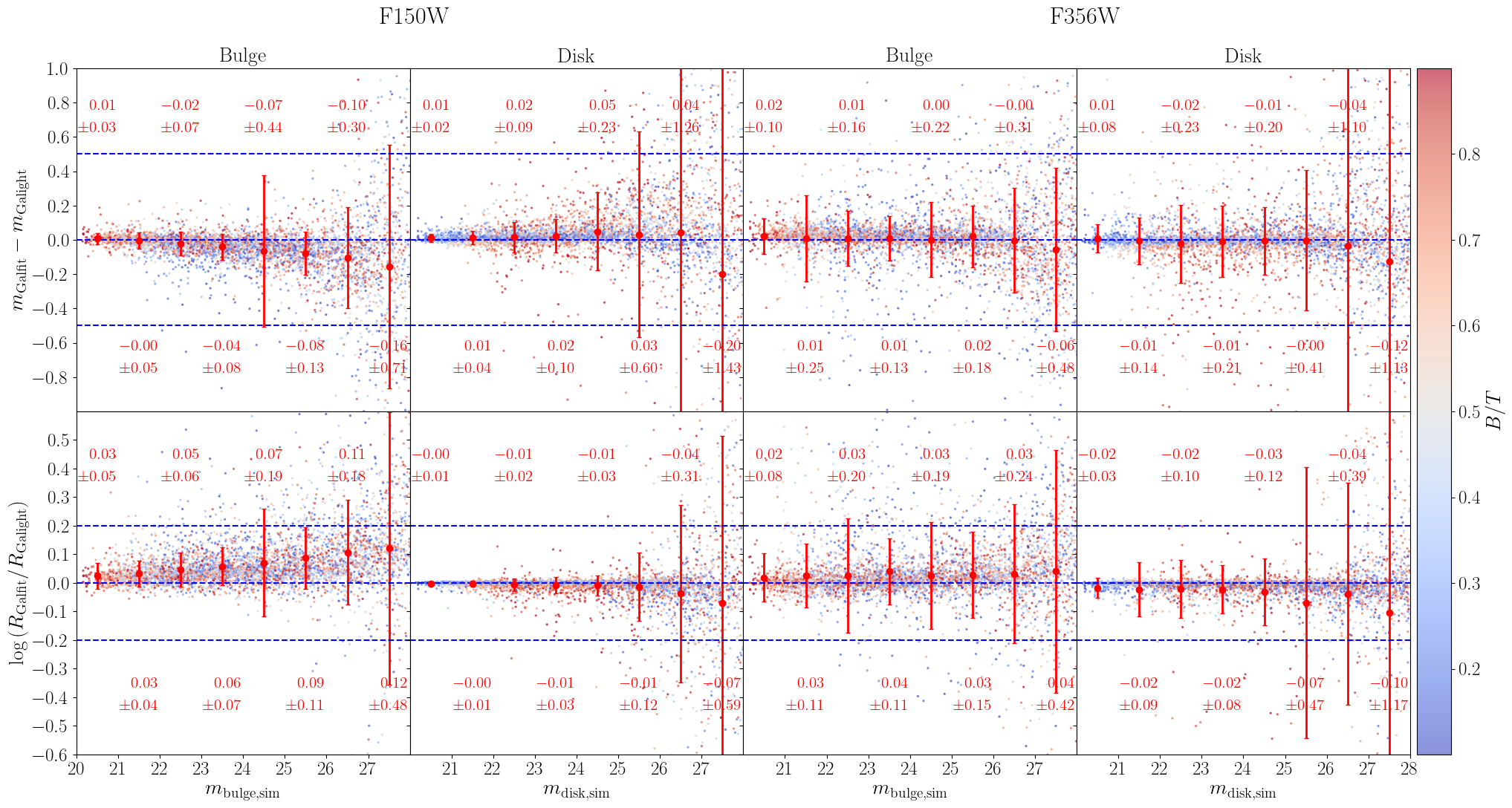}
    \caption{Comparison  between \texttt{galfit} and \galight\ for magnitude and effective radius of double \sersic\  model fitting. See caption of Figure~\ref{fig:5} for details.}
    \label{fig:10}
\end{figure*}
To confirm the robustness of this work, we compare the fitting results obtained from our modelling software, \galight, with those derived from \texttt{galfit}. We applied \texttt{galfit} to fit the same set of 5,000 simulated galaxies, using consistent fitting settings for both the single \sersic\ and double \sersic\ models. The results of this comparison are illustrated in Figure~\ref{fig:9}, which shows that for the single \sersic\ fitting, the inferred values are consistent between the two software packages for magnitude and size, with differences of less than 0.1 mag, and 0.1 dex (with \( \text{mag} <26 \)), respectively, for relatively brighter galaxies.

For the bulge+disk decomposition fitting, as shown in Figure~\ref{fig:10}, we also find that the inferred values for magnitude and effective radius \reff\ for both the bulge and disk components are consistent across \galight\ and \texttt{galfit}, with differences of less than 0.5 mag and 0.2 dex for brighter galaxies (with \( \text{mag} <26 \)) for both filters, respectively. This agreement reinforces the reliability of our fitting methods and suggests that both software packages can effectively model galaxy structures.

The consistent results obtained from both \galight\ and \texttt{galfit} indicate that our simulation results are robust and reliable. The close alignment in the inferred parameters across different fitting tools enhances our confidence in the validity of our findings and supports the use of these results as a systematic reference in future studies of galaxy photometry, morphology, and bulge+disk decomposition using high-quality data delivered by JWST.

\subsection{Core Components Before Complexity: Rationale for Bulge+Disk Focus}
\label{first_step}
Our findings demonstrate JWST's unprecedented capability to resolve galaxies' fundamental structures—bulges and disks—that encode critical information about galaxy assembly histories. These components represent the first-order morphological features observable at high redshifts, where bulges trace merger-driven growth~\citep[e.g.,][]{2009MNRAS.397..802H} and disks record gas accretion and in-situ star formation~\citep[e.g.,][]{2018MNRAS.474.3976G}. While substructures like bars, clumps, and spirals undoubtedly influence galaxy evolution, our focused analysis establishes essential baselines. At $z > 2$, even JWST struggles to resolve sub-kiloparsec features in most galaxies, making bulge/disk separation the primary achievable morphological measurement~\citep[e.g.,][]{van2023stellar}. By first quantifying systematic errors in this "minimum complexity" scenario (i.e., one \sersic\ $n=1$ disk plus one \sersic\ $n=4$ bulge), we provide the community with essential error budgets and selection criteria for interpreting high-$z$ structures.

While our current analysis establishes valuable baselines, it assumes smooth, symmetric bulge+disk systems. Real galaxies, especially at high redshift, often exhibit morphological irregularities such as clumps, off-centered features, and non-axisymmetric structures that are not captured in our models. In the literature, studies have shown that stellar bars, ubiquitous in disk galaxies, can significantly bias \sersic-based decomposition~\citep[e.g.,][]{2022MNRAS.513..693B, 2022ApJ...934...52B}. Furthermore, \citet{2019A&A...632A.128B} emphasized the fundamental limitations of \sersic\ modelling, particularly under the noisy, asymmetric conditions typical of high-redshift imaging.

Recent JWST/NIRCam results indicate that such features are common; for instance, \citet{2025MNRAS.537..402K} find that approximately 40\% of $z \sim 1$--$3$ galaxies exhibit clumpy substructures in near-infrared imaging. This indicates that idealized morphologies may not be representative for a large fraction of galaxies, motivating future simulations that more accurately reflect the structural complexity observed in the high-redshift Universe.

Our future work will extend our framework by incorporating realistic substructures (bars, spirals, clumps) into mock galaxies, enabling robust tests of decomposition reliability for JWST observations. Leveraging JWST's unprecedented resolution, we will systematically quantify how irregular morphologies in high-redshift galaxies differ from their low-redshift counterparts, which is a crucial test for understanding galaxy evolution.

\section{Summary}{\label{sec.6}}
Understanding galaxy morphology -- a key tracer of assembly history and dynamical evolution -- requires robust decomposition methods to disentangle the contributions of coexisting components like bulges and disks. With JWST's unprecedented resolution and sensitivity now studying galaxies at cosmic dawn, systematic validation of these techniques is critical. In this study, we carried out extensive simulation tests by generating mock JWST/NIRCam observations to test the accuracy and limitations of bulge+disk decompositions. Our simulations replicate JWST/NIRCam observing conditions from the CEERS survey, incorporating empirically PSF models and background noise level. Using these realistic observational conditions, we generated mock galaxies with systematically varied bulge+disk structural properties, enabling controlled tests of bulge+disk identification and quantitative assessment of parameter recovery accuracy. We systematically evaluated the fitting performance of both single \sersic\ and double \sersic\ (i.e., bulge+disk) models.

For a single \sersic\ case, we demonstrated that when the bulge and disk contribute comparable flux (e.g., $B/T\sim50\%$, Figure~\ref{fig:2}, bottom), single-component fits produce prominent residuals that clearly reveal the underlying composite structure. Conversely, in systems where one component dominates (e.g., $B/T\sim20\%$, Figure~\ref{fig:2}, top), the residuals become negligible, rendering the two-component nature indistinguishable through the fitting using the single \sersic\ model alone. We also quantified how accurately single \sersic\ parameters (total magnitude, effective radius, and \sersic\ index) capture the integrated properties of composite systems. Our results showed that single \sersic\ models can reliably recover total magnitudes (within 0.2 mag for galaxies as faint as 27 mag) and effective radius (within 0.2 dex for galaxies brighter than 27 mag) for both F150W and F356W. In Section~\ref{dis:n}, we further investigate the relationship between the inferred \sersic\ index ($n$) and the intrinsic B/T for our simulated galaxies. Our analysis reveals a robust correlation between $n$ and B/T across both F150W and F356W bands, demonstrating the single \sersic\ index's utility as a proxy for structural composition.

For the double \sersic\ case, our analysis demonstrates that the model robustly improves the residual for the former single \sersic\ fits -- when applied to systems with the sample comparable bulge and disk system (e.g., $B/T\sim50\%$, Figure~\ref{fig:4}), the two-component model eliminates systematic residuals and confirm the bulge+disk structure of this system. We then quantified the precision of double \sersic\ decompositions in measuring individual component properties. We found that both bulge and disk magnitudes are recovered within 0.5 mag for components brighter than 26 mag, with better accuracy for dominant components (e.g., bulge residuals of $-0.042\pm0.151$ in F150W for $B/T>60\%$). The effective radius are recovered with systematic uncertainties below 0.2 dex for bright components (mag$<26$), leveraging JWST's high resolution to mitigate degeneracies between compact bulges and extended disks. Notably, the size measurement accuracy in F150W outperforms F356W, due to its superior angular resolution, which better resolves small-scale structures. Additionally, the extended nature of disks leads to more robust size inferences compared to the more compact bulge component.

A key outcome of this study is the systematic quantification of biases and uncertainties in morphological parameters as a function of SNR. We establish empirical relations between SNR and the accuracy of recovered properties (including magnitudes and sizes) for both single and double \sersic\ models (see Figure~\ref{fig:6} and~\ref{fig:7}), serving as a practical framework to estimate measurement uncertainties in other JWST/NIRCam programs.

To determine whether galaxies are better described by single or double \sersic\ models, we performed model selection using the Bayesian Information Criterion (BIC). Our analysis reveals a strong preference for double \sersic\ fits in galaxies with sufficient SNR: we found that 90\% of galaxies brighter than 24.9 mag (F150W) or 26.2 mag (F356W) exhibited $\Delta$BIC$<-10$. For galaxies with comparable bulge and disk contributions ($30\% < B/T < 70\%$), the double \sersic\ model is strongly favored (90\% of them meet $\Delta$BIC$<-10$). These findings provide clear magnitude thresholds for future JWST studies, helping future studies to decide when to apply single vs. double-component models based on galaxy brightness and B/T ratios.

Our study provides the first empirical error budgets for JWST-based morphological analyses, establishing practical thresholds for future studies. These results will guide high-redshift research linking structures to formation mechanisms (e.g., merger-driven bulges vs. disk growth). Future work will address current limitations, such as the exclusion of pseudo-bulges and complex substructures (e.g., bar, spiral, and clumpy features), to further refine decomposition techniques for JWST's next-generation datasets.

\section*{Acknowledgements}
We sincerely thank Hassen for his valuable discussions and insightful suggestions. This work was supported by  the National Key Research and Development Program of China (No. 2024YFC2207400).

\section*{Data Availability}
The simulated data underlying this article will be shared on reasonable request to the first or corresponding author.



\bibliographystyle{mnras}
\bibliography{references} 

\begin{thebibliography}{}
\makeatletter
\relax
\def\mn@urlcharsother{\let\do\@makeother \do\$\do\&\do\#\do\^\do\_\do\%\do\~}
\def\mn@doi{\begingroup\mn@urlcharsother \@ifnextchar [ {\mn@doi@}
  {\mn@doi@[]}}
\def\mn@doi@[#1]#2{\def\@tempa{#1}\ifx\@tempa\@empty \href
  {http://dx.doi.org/#2} {doi:#2}\else \href {http://dx.doi.org/#2} {#1}\fi
  \endgroup}
\def\mn@eprint#1#2{\mn@eprint@#1:#2::\@nil}
\def\mn@eprint@arXiv#1{\href {http://arxiv.org/abs/#1} {{\tt arXiv:#1}}}
\def\mn@eprint@dblp#1{\href {http://dblp.uni-trier.de/rec/bibtex/#1.xml}
  {dblp:#1}}
\def\mn@eprint@#1:#2:#3:#4\@nil{\def\@tempa {#1}\def\@tempb {#2}\def\@tempc
  {#3}\ifx \@tempc \@empty \let \@tempc \@tempb \let \@tempb \@tempa \fi \ifx
  \@tempb \@empty \def\@tempb {arXiv}\fi \@ifundefined
  {mn@eprint@\@tempb}{\@tempb:\@tempc}{\expandafter \expandafter \csname
  mn@eprint@\@tempb\endcsname \expandafter{\@tempc}}}

\bibitem[\protect\citeauthoryear{{Athanassoula}}{{Athanassoula}}{2005}]{2005MNRAS.358.1477A}
{Athanassoula} E.,  2005, \mn@doi [\mnras] {10.1111/j.1365-2966.2005.08872.x},
  \href {https://ui.adsabs.harvard.edu/abs/2005MNRAS.358.1477A} {358, 1477}

\bibitem[\protect\citeauthoryear{{Bernardi}, {Meert}, {Vikram},
  {Huertas-Company}, {Mei}, {Shankar}  \& {Sheth}}{{Bernardi}
  et~al.}{2014}]{2014MNRAS.443..874B}
{Bernardi} M.,  {Meert} A.,  {Vikram} V.,  {Huertas-Company} M.,  {Mei} S.,
  {Shankar} F.,   {Sheth} R.~K.,  2014, \mn@doi [\mnras]
  {10.1093/mnras/stu1106}, \href
  {https://ui.adsabs.harvard.edu/abs/2014MNRAS.443..874B} {443, 874}

\bibitem[\protect\citeauthoryear{{Bi}, {Shlosman}  \& {Romano-D{\'\i}az}}{{Bi}
  et~al.}{2022a}]{2022MNRAS.513..693B}
{Bi} D.,  {Shlosman} I.,   {Romano-D{\'\i}az} E.,  2022a, \mn@doi [\mnras]
  {10.1093/mnras/stac363}, \href
  {https://ui.adsabs.harvard.edu/abs/2022MNRAS.513..693B} {513, 693}

\bibitem[\protect\citeauthoryear{{Bi}, {Shlosman}  \& {Romano-D{\'\i}az}}{{Bi}
  et~al.}{2022b}]{2022ApJ...934...52B}
{Bi} D.,  {Shlosman} I.,   {Romano-D{\'\i}az} E.,  2022b, \mn@doi [\apj]
  {10.3847/1538-4357/ac779b}, \href
  {https://ui.adsabs.harvard.edu/abs/2022ApJ...934...52B} {934, 52}

\bibitem[\protect\citeauthoryear{{Birrer} \& {Amara}}{{Birrer} \&
  {Amara}}{2018}]{2018PDU....22..189B}
{Birrer} S.,  {Amara} A.,  2018, \mn@doi [Physics of the Dark Universe]
  {10.1016/j.dark.2018.11.002}, \href
  {https://ui.adsabs.harvard.edu/abs/2018PDU....22..189B} {22, 189}

\bibitem[\protect\citeauthoryear{{Birrer} et~al.,}{{Birrer}
  et~al.}{2021}]{2021JOSS....6.3283B}
{Birrer} S.,  et~al., 2021, \mn@doi [The Journal of Open Source Software]
  {10.21105/joss.03283}, \href
  {https://ui.adsabs.harvard.edu/abs/2021JOSS....6.3283B} {6, 3283}

\bibitem[\protect\citeauthoryear{Blanchard et~al.,}{Blanchard
  et~al.}{2024}]{blanchard2024jwst}
Blanchard P.~K.,  et~al., 2024, Nature Astronomy, 8, 774

\bibitem[\protect\citeauthoryear{{Breda}, {Papaderos}, {Gomes}  \&
  {Amarantidis}}{{Breda} et~al.}{2019}]{2019A&A...632A.128B}
{Breda} I.,  {Papaderos} P.,  {Gomes} J.~M.,   {Amarantidis} S.,  2019, \mn@doi
  [\aap] {10.1051/0004-6361/201935144}, \href
  {https://ui.adsabs.harvard.edu/abs/2019A&A...632A.128B} {632, A128}

\bibitem[\protect\citeauthoryear{{Bruce} et~al.,}{{Bruce}
  et~al.}{2014}]{bruce2014}
{Bruce} V.~A.,  et~al., 2014, \mn@doi [\mnras] {10.1093/mnras/stu1478}, \href
  {https://ui.adsabs.harvard.edu/abs/2014MNRAS.444.1001B} {444, 1001}

\bibitem[\protect\citeauthoryear{Casura et~al.,}{Casura
  et~al.}{2022}]{casura2022galaxy}
Casura S.,  et~al., 2022, Monthly Notices of the Royal Astronomical Society,
  516, 942

\bibitem[\protect\citeauthoryear{{Conselice}}{{Conselice}}{2014}]{2014ARA&A..52..291C}
{Conselice} C.~J.,  2014, \mn@doi [\araa]
  {10.1146/annurev-astro-081913-040037}, \href
  {https://ui.adsabs.harvard.edu/abs/2014ARA&A..52..291C} {52, 291}

\bibitem[\protect\citeauthoryear{Ding et~al.,}{Ding et~al.}{2020}]{Ding_2020}
Ding X.,  et~al., 2020, \mn@doi [The Astrophysical Journal]
  {10.3847/1538-4357/ab5b90}, 888, 37

\bibitem[\protect\citeauthoryear{{Ding}, {Silverman}  \& {Onoue}}{{Ding}
  et~al.}{2022}]{2022ApJ...939L..28D}
{Ding} X.,  {Silverman} J.~D.,   {Onoue} M.,  2022, \mn@doi [\apjl]
  {10.3847/2041-8213/ac9c02}, \href
  {https://ui.adsabs.harvard.edu/abs/2022ApJ...939L..28D} {939, L28}

\bibitem[\protect\citeauthoryear{{Ding} et~al.,}{{Ding}
  et~al.}{2023}]{2023Natur.621...51D}
{Ding} X.,  et~al., 2023, \mn@doi [\nat] {10.1038/s41586-023-06345-5}, \href
  {https://ui.adsabs.harvard.edu/abs/2023Natur.621...51D} {621, 51}

\bibitem[\protect\citeauthoryear{Du, Ho, Debattista, Pillepich, Nelson,
  Hernquist  \& Weinberger}{Du et~al.}{2021}]{du2021evolutionary}
Du M.,  Ho L.~C.,  Debattista V.~P.,  Pillepich A.,  Nelson D.,  Hernquist L.,
   Weinberger R.,  2021, The Astrophysical Journal, 919, 135

\bibitem[\protect\citeauthoryear{{Ferreira} et~al.,}{{Ferreira}
  et~al.}{2022}]{2022ApJ...938L...2F}
{Ferreira} L.,  et~al., 2022, \mn@doi [\apjl] {10.3847/2041-8213/ac947c}, \href
  {https://ui.adsabs.harvard.edu/abs/2022ApJ...938L...2F} {938, L2}

\bibitem[\protect\citeauthoryear{Finkelstein, Pirzkal, Malhotra  \&
  Rhoads}{Finkelstein et~al.}{2017}]{Finkelstein2017}
Finkelstein S.~L.,  Pirzkal N.,  Malhotra S.,   Rhoads J.~E.,  2017, \mn@doi
  [The Astrophysical Journal] {10.3847/1538-4357/aa6c80}, 839, 99

\bibitem[\protect\citeauthoryear{Finkelstein, Bagley, Ferguson
  et~al.}{Finkelstein et~al.}{2022}]{Finkelstein2022}
Finkelstein S.~L.,  Bagley M.,  Ferguson H.~C.,   et~al., 2022, \mn@doi [The
  Astrophysical Journal] {10.3847/1538-4357/ac84b1}, 936, 80

\bibitem[\protect\citeauthoryear{{Fisher} \& {Drory}}{{Fisher} \&
  {Drory}}{2008}]{fisher2008}
{Fisher} D.~B.,  {Drory} N.,  2008, \mn@doi [\aj]
  {10.1088/0004-6256/136/2/773}, \href
  {https://ui.adsabs.harvard.edu/abs/2008AJ....136..773F} {136, 773}

\bibitem[\protect\citeauthoryear{{Freeman}}{{Freeman}}{1970}]{1970ApJ...160..811F}
{Freeman} K.~C.,  1970, \mn@doi [\apj] {10.1086/150474}, \href
  {https://ui.adsabs.harvard.edu/abs/1970ApJ...160..811F} {160, 811}

\bibitem[\protect\citeauthoryear{{Gadotti}}{{Gadotti}}{2009a}]{2009MNRAS.393.1531G}
{Gadotti} D.~A.,  2009a, \mn@doi [\mnras] {10.1111/j.1365-2966.2008.14257.x},
  \href {https://ui.adsabs.harvard.edu/abs/2009MNRAS.393.1531G} {393, 1531}

\bibitem[\protect\citeauthoryear{Gadotti}{Gadotti}{2009b}]{gadotti2009}
Gadotti D.~A.,  2009b, Monthly Notices of the Royal Astronomical Society, 393,
  1531

\bibitem[\protect\citeauthoryear{{Genel} et~al.,}{{Genel}
  et~al.}{2018}]{2018MNRAS.474.3976G}
{Genel} S.,  et~al., 2018, \mn@doi [\mnras] {10.1093/mnras/stx3078}, \href
  {https://ui.adsabs.harvard.edu/abs/2018MNRAS.474.3976G} {474, 3976}

\bibitem[\protect\citeauthoryear{{Giavalisco}, {Steidel}  \&
  {Macchetto}}{{Giavalisco} et~al.}{1996}]{1996ApJ...470..189G}
{Giavalisco} M.,  {Steidel} C.~C.,   {Macchetto} F.~D.,  1996, \mn@doi [\apj]
  {10.1086/177859}, \href
  {https://ui.adsabs.harvard.edu/abs/1996ApJ...470..189G} {470, 189}

\bibitem[\protect\citeauthoryear{{H{\"a}u{\ss}ler} et~al.,}{{H{\"a}u{\ss}ler}
  et~al.}{2013}]{2013MNRAS.430..330H}
{H{\"a}u{\ss}ler} B.,  et~al., 2013, \mn@doi [\mnras] {10.1093/mnras/sts633},
  \href {https://ui.adsabs.harvard.edu/abs/2013MNRAS.430..330H} {430, 330}

\bibitem[\protect\citeauthoryear{{Hopkins} et~al.,}{{Hopkins}
  et~al.}{2009}]{2009MNRAS.397..802H}
{Hopkins} P.~F.,  et~al., 2009, \mn@doi [\mnras]
  {10.1111/j.1365-2966.2009.14983.x}, \href
  {https://ui.adsabs.harvard.edu/abs/2009MNRAS.397..802H} {397, 802}

\bibitem[\protect\citeauthoryear{{Jogee} et~al.,}{{Jogee}
  et~al.}{2004}]{2004ApJ...615L.105J}
{Jogee} S.,  et~al., 2004, \mn@doi [\apjl] {10.1086/426138}, \href
  {https://ui.adsabs.harvard.edu/abs/2004ApJ...615L.105J} {615, L105}

\bibitem[\protect\citeauthoryear{{Kalita}, {Silverman}, {Daddi}, {Mercier},
  {Ho}  \& {Ding}}{{Kalita} et~al.}{2025}]{2025MNRAS.537..402K}
{Kalita} B.~S.,  {Silverman} J.~D.,  {Daddi} E.,  {Mercier} W.,  {Ho} L.~C.,
  {Ding} X.,  2025, \mn@doi [\mnras] {10.1093/mnras/staf031}, \href
  {https://ui.adsabs.harvard.edu/abs/2025MNRAS.537..402K} {537, 402}

\bibitem[\protect\citeauthoryear{Kelly et~al.}{Kelly et~al.}{2023}]{Kelly2023}
Kelly P.,  et~al., 2023, The Astrophysical Journal

\bibitem[\protect\citeauthoryear{Kennedy \& Eberhart}{Kennedy \&
  Eberhart}{1995}]{Kennedy1995}
Kennedy J.,  Eberhart R.,  1995, in Proceedings of the IEEE International
  Conference on Neural Networks. pp 1942--1945

\bibitem[\protect\citeauthoryear{Kormendy}{Kormendy}{2013}]{kormendy2013secular}
Kormendy J.,  2013, arXiv preprint arXiv:1311.2609

\bibitem[\protect\citeauthoryear{{Kormendy} \& {Kennicutt}}{{Kormendy} \&
  {Kennicutt}}{2004}]{2004ARA&A..42..603K}
{Kormendy} J.,  {Kennicutt} Jr. R.~C.,  2004, \mn@doi [\araa]
  {10.1146/annurev.astro.42.053102.134024}, \href
  {https://ui.adsabs.harvard.edu/abs/2004ARA&A..42..603K} {42, 603}

\bibitem[\protect\citeauthoryear{{Kormendy}, {Fisher}, {Cornell}  \&
  {Bender}}{{Kormendy} et~al.}{2009}]{2009ApJS..182..216K}
{Kormendy} J.,  {Fisher} D.~B.,  {Cornell} M.~E.,   {Bender} R.,  2009, \mn@doi
  [\apjs] {10.1088/0067-0049/182/1/216}, \href
  {https://ui.adsabs.harvard.edu/abs/2009ApJS..182..216K} {182, 216}

\bibitem[\protect\citeauthoryear{LaChance, Croft, Ni, Chen, Matteo  \&
  Bird}{LaChance et~al.}{2025}]{LaChance_2025}
LaChance P.,  Croft R.,  Ni Y.,  Chen N.,  Matteo T.~D.,   Bird S.,  2025,
  \mn@doi [The Open Journal of Astrophysics] {10.33232/001c.129991}, 8

\bibitem[\protect\citeauthoryear{{Lackner} \& {Gunn}}{{Lackner} \&
  {Gunn}}{2012}]{2012MNRAS.421.2277L}
{Lackner} C.~N.,  {Gunn} J.~E.,  2012, \mn@doi [\mnras]
  {10.1111/j.1365-2966.2012.20450.x}, \href
  {https://ui.adsabs.harvard.edu/abs/2012MNRAS.421.2277L} {421, 2277}

\bibitem[\protect\citeauthoryear{Lambas, Maddox  \& Loveday}{Lambas
  et~al.}{1992}]{10.1093/mnras/258.2.404}
Lambas D.~G.,  Maddox S.~J.,   Loveday J.,  1992, \mn@doi [Monthly Notices of
  the Royal Astronomical Society] {10.1093/mnras/258.2.404}, 258, 404

\bibitem[\protect\citeauthoryear{{Lang} et~al.,}{{Lang}
  et~al.}{2014}]{2014ApJ...788...11L}
{Lang} P.,  et~al., 2014, \mn@doi [\apj] {10.1088/0004-637X/788/1/11}, \href
  {https://ui.adsabs.harvard.edu/abs/2014ApJ...788...11L} {788, 11}

\bibitem[\protect\citeauthoryear{{Le Bail} et~al.,}{{Le Bail}
  et~al.}{2024}]{lebail2024}
{Le Bail} A.,  et~al., 2024, \mn@doi [\aap] {10.1051/0004-6361/202347465},
  \href {https://ui.adsabs.harvard.edu/abs/2024A&A...688A..53L} {688, A53}

\bibitem[\protect\citeauthoryear{{Meert}, {Vikram}  \& {Bernardi}}{{Meert}
  et~al.}{2013a}]{2013MNRAS.433.1344M}
{Meert} A.,  {Vikram} V.,   {Bernardi} M.,  2013a, \mn@doi [\mnras]
  {10.1093/mnras/stt822}, \href
  {https://ui.adsabs.harvard.edu/abs/2013MNRAS.433.1344M} {433, 1344}

\bibitem[\protect\citeauthoryear{Meert, Vikram  \& Bernardi}{Meert
  et~al.}{2013b}]{meert2013simulations}
Meert A.,  Vikram V.,   Bernardi M.,  2013b, Monthly Notices of the Royal
  Astronomical Society, 433, 1344

\bibitem[\protect\citeauthoryear{Miller et~al.,}{Miller
  et~al.}{2024}]{miller2024jwstuncoversopticalsize}
Miller T.~B.,  et~al., 2024, JWST UNCOVERs the Optical Size - Stellar Mass
  Relation at $4<z<8$: Rapid Growth in the Sizes of Low Mass Galaxies in the
  First Billion Years of the Universe (\mn@eprint {arXiv} {2412.06957}), \url
  {https://arxiv.org/abs/2412.06957}

\bibitem[\protect\citeauthoryear{Mo, Van~den Bosch  \& White}{Mo
  et~al.}{2010}]{mo2010galaxy}
Mo H.,  Van~den Bosch F.,   White S.,  2010, Galaxy formation and evolution.
Cambridge University Press

\bibitem[\protect\citeauthoryear{Montes \& Trujillo}{Montes \&
  Trujillo}{2022}]{montes2022new}
Montes M.,  Trujillo I.,  2022, The Astrophysical Journal Letters, 940, L51

\bibitem[\protect\citeauthoryear{Mosleh, Williams  \& Franx}{Mosleh
  et~al.}{2013}]{mosleh2013robustness}
Mosleh M.,  Williams R.~J.,   Franx M.,  2013, The Astrophysical Journal, 777,
  117

\bibitem[\protect\citeauthoryear{Newman, Ellis  \& Bundy}{Newman
  et~al.}{2012}]{Newman2012}
Newman A.~B.,  Ellis R.~S.,   Bundy K. e.~a.,  2012, \mn@doi [ApJ]
  {10.1088/0004-637X/746/2/162}, 746, 162

\bibitem[\protect\citeauthoryear{{Peng}, {Ho}, {Impey}  \& {Rix}}{{Peng}
  et~al.}{2002}]{peng2002detailed}
{Peng} C.~Y.,  {Ho} L.~C.,  {Impey} C.~D.,   {Rix} H.-W.,  2002, \mn@doi [\aj]
  {10.1086/340952}, \href
  {https://ui.adsabs.harvard.edu/abs/2002AJ....124..266P} {124, 266}

\bibitem[\protect\citeauthoryear{Peng, Ho, Impey  \& Rix}{Peng
  et~al.}{2010}]{peng2010detailed}
Peng C.~Y.,  Ho L.~C.,  Impey C.~D.,   Rix H.-W.,  2010, The Astronomical
  Journal, 139, 2097

\bibitem[\protect\citeauthoryear{{Rieke} et~al.,}{{Rieke}
  et~al.}{2023}]{2023PASP..135b8001R}
{Rieke} M.~J.,  et~al., 2023, \mn@doi [\pasp] {10.1088/1538-3873/acac53}, \href
  {https://ui.adsabs.harvard.edu/abs/2023PASP..135b8001R} {135, 028001}

\bibitem[\protect\citeauthoryear{Rigby, Perrin, McElwain  et~al.}{Rigby
  et~al.}{2022}]{Rigby2022}
Rigby J.~R.,  Perrin M.~D.,  McElwain M.~W.,   et~al., 2022, \mn@doi
  [Publications of the Astronomical Society of the Pacific]
  {10.1088/1538-3873/ac8a3a}, 134, 073001

\bibitem[\protect\citeauthoryear{{Sandage}}{{Sandage}}{2005}]{2005ARA&A..43..581S}
{Sandage} A.,  2005, \mn@doi [\araa] {10.1146/annurev.astro.43.112904.104839},
  \href {https://ui.adsabs.harvard.edu/abs/2005ARA&A..43..581S} {43, 581}

\bibitem[\protect\citeauthoryear{{Simard}, {Mendel}, {Patton}, {Ellison}  \&
  {McConnachie}}{{Simard} et~al.}{2011a}]{2011ApJS..196...11S}
{Simard} L.,  {Mendel} J.~T.,  {Patton} D.~R.,  {Ellison} S.~L.,
  {McConnachie} A.~W.,  2011a, \mn@doi [\apjs] {10.1088/0067-0049/196/1/11},
  \href {https://ui.adsabs.harvard.edu/abs/2011ApJS..196...11S} {196, 11}

\bibitem[\protect\citeauthoryear{Simard, Trevor~Mendel, Patton, Ellison  \&
  McConnachie}{Simard et~al.}{2011b}]{Simard_2011}
Simard L.,  Trevor~Mendel J.,  Patton D.~R.,  Ellison S.~L.,   McConnachie
  A.~W.,  2011b, \mn@doi [The Astrophysical Journal Supplement Series]
  {10.1088/0067-0049/196/1/11}, 196, 11

\bibitem[\protect\citeauthoryear{Sun, Ho, Zhuang, Ma, Chen  \& Li}{Sun
  et~al.}{2024}]{sun2024structure}
Sun W.,  Ho L.~C.,  Zhuang M.-Y.,  Ma C.,  Chen C.,   Li R.,  2024, The
  Astrophysical Journal, 960, 104

\bibitem[\protect\citeauthoryear{Sánchez-Janssen \& Gadotti}{Sánchez-Janssen
  \& Gadotti}{2016}]{Sanchez2016}
Sánchez-Janssen R.,  Gadotti D.~A.,  2016, \mn@doi [MNRAS]
  {10.1093/mnras/stv2896}, 456, 3378

\bibitem[\protect\citeauthoryear{Sérsic}{Sérsic}{1963}]{sersic1963}
Sérsic J.~L.,  1963, Boletin de la Asociacion Argentina de Astronomia, 6, 41

\bibitem[\protect\citeauthoryear{{Tabor}, {Merrifield}, {Arag{\'o}n-Salamanca},
  {Cappellari}, {Bamford}  \& {Johnston}}{{Tabor}
  et~al.}{2017}]{2017MNRAS.466.2024T}
{Tabor} M.,  {Merrifield} M.,  {Arag{\'o}n-Salamanca} A.,  {Cappellari} M.,
  {Bamford} S.~P.,   {Johnston} E.,  2017, \mn@doi [\mnras]
  {10.1093/mnras/stw3183}, \href
  {https://ui.adsabs.harvard.edu/abs/2017MNRAS.466.2024T} {466, 2024}

\bibitem[\protect\citeauthoryear{{Treu} et~al.,}{{Treu}
  et~al.}{2023}]{Treu2023}
{Treu} T.,  et~al., 2023, \mn@doi [\apjl] {10.3847/2041-8213/ac9283}, \href
  {https://ui.adsabs.harvard.edu/abs/2023ApJ...942L..28T} {942, L28}

\bibitem[\protect\citeauthoryear{{Trujillo}, {Conselice}, {Bundy}, {Cooper},
  {Eisenhardt}  \& {Ellis}}{{Trujillo} et~al.}{2007}]{2007MNRAS.382..109T}
{Trujillo} I.,  {Conselice} C.~J.,  {Bundy} K.,  {Cooper} M.~C.,  {Eisenhardt}
  P.,   {Ellis} R.~S.,  2007, \mn@doi [\mnras]
  {10.1111/j.1365-2966.2007.12388.x}, \href
  {https://ui.adsabs.harvard.edu/abs/2007MNRAS.382..109T} {382, 109}

\bibitem[\protect\citeauthoryear{{Yang} et~al.,}{{Yang}
  et~al.}{2022}]{Yang2022}
{Yang} L.,  et~al., 2022, \mn@doi [\apjl] {10.3847/2041-8213/ac8803}, \href
  {https://ui.adsabs.harvard.edu/abs/2022ApJ...938L..17Y} {938, L17}

\bibitem[\protect\citeauthoryear{{Yang} et~al.,}{{Yang}
  et~al.}{2025}]{Yang2025}
{Yang} L.,  et~al., 2025, arXiv e-prints, \href
  {https://ui.adsabs.harvard.edu/abs/2025arXiv250407185Y} {p. arXiv:2504.07185}

\bibitem[\protect\citeauthoryear{{de Vaucouleurs}}{{de
  Vaucouleurs}}{1948a}]{1948AnAp...11..247D}
{de Vaucouleurs} G.,  1948a, Annales d'Astrophysique, \href
  {https://ui.adsabs.harvard.edu/abs/1948AnAp...11..247D} {11, 247}

\bibitem[\protect\citeauthoryear{de
  Vaucouleurs}{de~Vaucouleurs}{1948b}]{de1948recherches}
de Vaucouleurs G.,  1948b, Annales d'Astrophysique, Vol. 11, p. 247, 11, 247

\bibitem[\protect\citeauthoryear{van~der Kruit \& Freeman}{van~der Kruit \&
  Freeman}{2011}]{van_der_Kruit_2011}
van~der Kruit P.,  Freeman K.,  2011, \mn@doi [Annual Review of Astronomy and
  Astrophysics] {10.1146/annurev-astro-083109-153241}, 49, 301–371

\bibitem[\protect\citeauthoryear{{van der Wel} et~al.,}{{van der Wel}
  et~al.}{2012}]{2012ApJS..203...24V}
{van der Wel} A.,  et~al., 2012, \mn@doi [\apjs] {10.1088/0067-0049/203/2/24},
  \href {https://ui.adsabs.harvard.edu/abs/2012ApJS..203...24V} {203, 24}

\bibitem[\protect\citeauthoryear{van~der Wel, Franx  \& van Dokkum}{van~der Wel
  et~al.}{2014}]{vanderWel2014}
van~der Wel A.,  Franx M.,   van Dokkum P. G. e.~a.,  2014, \mn@doi [ApJ]
  {10.1088/0004-637X/788/1/28}, 788, 28

\bibitem[\protect\citeauthoryear{van~der Wel et~al.,}{van~der Wel
  et~al.}{2023}]{van2023stellar}
van~der Wel A.,  et~al., 2023, The Astrophysical Journal, 960, 53

\makeatother
\end{thebibliography}







\bsp	
\label{lastpage}
\end{document}